# Establishing correction solutions for Scanning Laser Doppler Vibrometer measurements affected by sensor head vibration

Ben J. Halkon[1] and Steve J. Rothberg[2]

[1]Centre for Audio, Acoustics & Vibration, Faculty of Engineering & IT, University of Technology Sydney, Ultimo, NSW 2007, Australia

[2]Wolfson School of Mechanical, Electrical and Manufacturing Engineering, Loughborough University, Loughborough, Leicestershire, LE11 3TU, UK.

Corresponding author: Ben Halkon

e-mail: Benjamin.Halkon@uts.edu.au,

tel: +61 (0) 2 9514 9442

# Establishing correction solutions for Scanning Laser Doppler Vibrometer measurements affected by sensor head vibration

Ben J. Halkon[1] and Steve J. Rothberg[2]

[1]Centre for Audio, Acoustics & Vibration, Faculty of Engineering & IT, University of Technology Sydney, Ultimo, NSW 2007, Australia

[2]Wolfson School of Mechanical, Electrical and Manufacturing Engineering, Loughborough University, Loughborough, Leicestershire, LE11 3TU, UK.

## Abstract

Scanning Laser Doppler Vibrometer (SLDV) measurements are affected by sensor head vibrations as if they are vibrations of the target surface itself. This paper presents practical correction schemes to solve this important problem. The study begins with a theoretical analysis, for arbitrary vibration and any scanning configuration, which shows that the only measurement required is of the vibration velocity at the incident point on the final steering mirror in the direction of the outgoing laser beam and this underpins the two correction options investigated. Correction sensor location is critical; the first scheme uses an accelerometer pair located on the SLDV front panel, either side of the emitted laser beam, while the second uses a single accelerometer located along the optical axis behind the final steering mirror.

Initial experiments with a vibrating sensor head and stationary target confirmed the sensitivity to sensor head vibration together with the effectiveness of the correction schemes which reduced overall error by 17 dB (accelerometer pair) and 27 dB (single accelerometer). In extensive further tests with both sensor head and target vibration, conducted across a range of scan angles, the correction schemes reduced error by typically 14 dB (accelerometer pair) and 20 dB (single accelerometer). RMS phase error was also up to 30% lower for the single accelerometer option, confirming it as the preferred option. The theory suggests a geometrical weighting of the correction measurements and this provides a small additional improvement.

Since the direction of the outgoing laser beam and its incident point on the final steering mirror both change as the mirrors scan the laser beam, the use of fixed axis correction transducers mounted in fixed locations makes the correction imperfect. The associated errors are estimated and expected to be generally small, and the theoretical basis for an enhanced three-axis correction is presented.

# 1 Introduction

The Scanning Laser Doppler Vibrometer (SLDV) has become widely deployed in many application domains as a convenient, reliable and accurate whole-field vibration measurement system [1]. Myriad applications have been conceived, researched, developed and largely perfected with many now readily accepted alongside those using more conventional sensors such as accelerometers or strain gauges. Indeed, in many cases, SLDV-based measurement campaigns offer significantly enhanced insights due to their inherent non-invasiveness, measurement of the optimum vibration parameter (velocity), higher spatial resolution, and wider frequency and dynamic ranges, with further practical benefits from remote operation and speed of set-up.

This paper addresses one of the remaining challenges to overcome, which is the sensitivity of the measurement to the SLDV's own vibration because LDVs, whether scanning or fixed beam, make a *relative* vibration velocity measurement. Vibration of the LDV body or of any beam steering optical elements, either within or external to the instrument, introduce additional Doppler shifts to the laser beam which add erroneously to the measured velocity and are indistinguishable from the desired measurement of target surface velocity. In some applications, these erroneous contributions might be low level or in a frequency range that allows them to be differentiated from the intended measurement and disregarded. For example, tripod mounting of the instrument, which is routine with SLDV, might ensure instrument vibrations are confined to a low frequency range below that of interest for target vibration. In general, however, correction is the only reliable method of eliminating these erroneous velocity components and this paper sets out the theoretical basis and practical implementation of schemes to solve this problem.

Employing common principles, earlier studies have demonstrated measurement correction for instrument vibration [2] and for steering optic vibration [3]. The subject of these studies was the standard (fixed/single beam) LDV, with a comprehensive mathematical approach used to determine the total velocity measured for arbitrary target *and* instrument/steering optic vibrations. The output from this totally general, vector-based method provided the theoretical basis for the practical compensation scheme which has been shown experimentally to offer (near) complete correction of the measured velocity for typical real-world vibration levels [2]-[4]. Recent similar campaigns have been reported, some making use of and extending these recommendations [5], and others employing alternative approaches with some success [6].

This comprehensive study builds on these earlier studies and includes a confirmation of the measurement sensitivity to sensor head vibration, the theoretical basis of the proposed correction schemes for arbitrary instrument vibration, and real-world relevant laboratory tests to confirm their effectiveness. To the authors' knowledge, this is the first published presentation of practical schemes to correct errors associated with vibrations of the sensor head for Scanning LDVs.

## 2 Theoretical basis for the correction of SLDV sensor head vibration

### 2.1 General considerations

It has been shown in a variety of circumstances [2], [3], [7] how the Doppler shifts associated with the vibrations of the instrument itself and of beam steering devices contribute additively to the measurement being made. Consequently, in a system such as the Scanning LDV with multiple beam steering devices, compensations for each of these additive Doppler shifts can also be summed to correct measurements. Summing compensations in this way, however, represents a significant measurement burden from both cost and practicality perspectives. The SLDV head used in this work, shown schematically in Figure 1, would require one or two accelerometers mounted on the LDV body [2], [4] and at least one accelerometer mounted on each of the steering mirrors [3]. The burden is compounded by the problem of mounting a contacting transducer on a scanning mirror which would severely challenge the stability of the mirror galvanometer dynamics. The LDV body and the scanning mirrors move together as a rigid body, however, and this paper will show theoretically and confirm experimentally how this enables a much more practical solution to the vibration compensation challenge.

[INSERT: Figure 1. Experimental arrangement schematic showing laser beam path and local coordinate system.]

The test SLDV used the same pair of scanning mirrors, associated electronics and optical layout as the Polytec Scanning Vibrometer PSV 300 [8] but instead incorporated a Polytec Compact Laser Vibrometer NLV-2500-5 [9] as the measurement LDV sensor head. The custom system affords good access to the preferred correction measurement locations, as can be seen in Figure 2 which shows several views of the internal optical arrangement corresponding to Figure 1. However, it is important to emphasise that the proposed measurement correction scheme is sufficiently simple and practical that it can be readily implemented in *any* system, commercial or otherwise.

[INSERT: Figure 2. Experimental arrangement *physical set-up*; a) LDV body mounted to bespoke SLDV assembly, b) top view with SLDV cover removed showing 'AccR', 'AccFL' and 'AccFR', and 'AccTar' with laser beam path super-imposed, c) side view showing SLDV assembly front panel with 'AccFL' & 'AccFR' and shaker with laser beam path super-imposed and d) close-up of vibrating target with 'AccTar'.]

### 2.2 Mathematical determination of the required correction

Figure 1 shows the instrument configuration in use with each beam orientation described by unit vectors. Following the path of the laser beam from the LDV sensor head, the first steering mirror is fixed, while the second and third mirrors are able to rotate to reorient, or scan, the laser beam. The laser beam exits the LDV body with direction unit vector $\hat{b}_1$ and passes through reference point O. The beam is then deflected from the fixed steering mirror at point A on the mirror surface, changing its direction to $\hat{b}_2$. Reflection from (fixed) point B on the 'y-scan' steering mirror changes the beam orientation to $\hat{b}_3$ and then reflection from (variable) point C on the 'x-scan' steering mirror redirects the beam to its final orientation, $\hat{b}_4$.

Following the vector-based approach [7], the total measured velocity, $U_m$, can be written as the sum of the velocities associated with the Doppler shifts resulting from LDV body vibration [2], the direction change at each mirror surface [3] and the desired measurement, $(\hat{b}_4 \cdot \vec{V_{T'}})$, of target vibration:

$$U_m = \hat{b}_1 \cdot \vec{V_O} + (\hat{b}_2 - \hat{b}_1) \cdot \vec{V_A} + (\hat{b}_3 - \hat{b}_2) \cdot \vec{V_B} + (\hat{b}_4 - \hat{b}_3) \cdot \vec{V_C} - \hat{b}_4 \cdot \vec{V_{T'}} \qquad (1)$$

With point O as the arbitrarily chosen reference point, the velocity vectors at each of the points A, B and C, described by position vectors $\vec{r_A}$, $\vec{r_B}$ and $\vec{r_C}$, can be written as follows:

$$\vec{V_A} = \vec{V_O} + \vec{r_A} \times \vec{\omega} \qquad (2a)$$

$$\vec{V_B} = \vec{V_O} + \vec{r_B} \times \vec{\omega} \qquad (2b)$$

$$\vec{V_C} = \vec{V_O} + \vec{r_C} \times \vec{\omega} \qquad (2c)$$

where $\vec{\omega}$ is the arbitrary angular velocity around the reference point. Equations (2a-c) make explicit the special relationship between the velocities at the incident points on the steering mirrors which are united by the reference velocity $\vec{V_O}$ and the angular velocity $\vec{\omega}$ under the reasonable assumption (based on consideration of the frequency range in which instrument vibration is likely in practice) that the SLDV instrument responds to ambient vibration as a rigid body.

Equation (1) can be reorganised to show the influence of the individual beam orientations:

$$U_m = \hat{b}_1 \cdot (\vec{V_O} - \vec{V_A}) + \hat{b}_2 \cdot (\vec{V_A} - \vec{V_B}) + \hat{b}_3 \cdot (\vec{V_B} - \vec{V_C}) + \hat{b}_4 \cdot (\vec{V_C} - \vec{V_{T'}}) \qquad (3a)$$

and then re-written by substituting in equations (2a-c):

$$U_m = -\hat{b}_1 \cdot (\vec{r_A} \times \vec{\omega}) + \hat{b}_2 \cdot ((\vec{r_A} - \vec{r_B}) \times \vec{\omega}) + \hat{b}_3 \cdot ((\vec{r_B} - \vec{r_C}) \times \vec{\omega}) + \hat{b}_4 \cdot (\vec{V_C} - \vec{V_{T'}}) \qquad (3b)$$

In the first triple scalar product, vectors $\hat{b}_1$ and $\vec{r_A}$ are by definition parallel so the triple scalar product, which can be re-ordered as $\vec{\omega} \cdot (\hat{b}_1 \times \vec{r_A})$, evaluates to zero. For the second and third triple scalar products, note how the vector differences $(\vec{r_A} - \vec{r_B})$ and $(\vec{r_B} - \vec{r_C})$ are each equal to vectors with direction equal to the associated laser beam orientations, $\hat{b}_2$ and $\hat{b}_3$ respectively. These triple scalar products also, therefore, evaluate to zero for the same reason as the first one. Consequently, the measured velocity is simply:

$$U_m = \hat{b}_4 \cdot (\vec{V_C} - \vec{V_{T'}}) \qquad (3c)$$

Equation (3c) represents the first significant outcome in this paper, for several reasons as follows. At the most general level, the analysis made no restrictive assumptions about the instrument vibration or the steering mirror orientations, all were entirely arbitrary. The analysis can also be generalised to any number of beam deflections since each will simply add another triple scalar product to equation (3b) that will evaluate to zero in the manner described above. Consequently, equation (3c) holds for arbitrary vibration of any SLDV head, provided the assumption of rigid body motion of LDV body and mirrors holds. Ultimately and most importantly because of the practical implication, it means that the compensation requires only a measurement of $(\hat{b}_4 . \overrightarrow{V_C})$ followed by subtraction from $U_m$ and so, at the very least, the correction measurement burden has been significantly reduced from multiple measurements on the LDV body and at each steering mirror to measurement of a single velocity component.

*2.3    Practical determination of the required correction*

However, while the measurement required for compensation might look much simpler now, it is not entirely straightforward to achieve practically. Firstly, the point C varies according to the orientation of any beam steering optics that precede this final steering mirror. In the setup used here [8], a beam deflection of 12 degrees by the first (*y*-scan) mirror causes the beam to move about 7 mm along the axis of the final (*x*-scan) mirror. Secondly, if a transducer is to be fixed to the mirror, as in previous steering mirror vibration compensation work [3], this time the required orientation is in the direction of the outgoing laser beam and not in the more convenient (from the perspective of mounting an accelerometer on the back of the mirror) mirror normal direction. Finally, the difficulty here is compounded by the beam orientation being necessarily variable in Scanning LDV.

The practical locations at which to locate transducers for compensating measurements follow the principles adopted in previous work [2]-[4]. In those applications, only a single axis measurement was required for theoretically perfect compensation. The essential principle is that the compensating measurement must be co-linear with the laser beam itself. The same principle applies in the SLDV application but the practicality associated with maintaining a measurement that is aligned with a scanning laser beam is problematic. This paper will first explore quantitively the viability of compensation with measurements that are perfect only when the beam is in its 'zero' position. This analysis is then developed to a more comprehensive proposal where all three components of vibration are measured for full compensation.

Practically, compensation options are limited to measurements at locations remote from C and in fixed directions. Following earlier successful approaches, the options available are location of a pair of transducers on the front face of the instrument equi-spaced either side of the line of the undeflected laser beam [2] or location of a single transducer in line with the undeflected laser beam [4] but at the rear of the final scanning mirror fixed to some convenient part of the mirror housing (not the mirror itself).

With the scanning mirrors in their 'zero' positions and the LDV optical axis perfectly aligned with that of the scanning system, the laser beam will be incident on the mirror at point C0, defined by a position vector $\vec{r_{C0}}$. The velocity at this point, $\vec{V_{C0}}$, is:

$$\vec{V_{C0}} = \vec{V_0} + \vec{r_{C0}} \times \vec{\omega} \qquad (4)$$

For compensating measurements using the accelerometer pair on the front face of the instrument (left and right, shown as 'AccFL' and 'AccFR' in Figure 1), the position vectors, $\vec{r_{FL}}$ and $\vec{r_{FR}}$, are:

$$\vec{r_{FL}} = \vec{r_{C0}} + r_{C0/Fx}\hat{x}_{LDV} + r_{C0/Fy}\hat{y}_{LDV} + r_{C0/Fz}\hat{z}_{LDV} \qquad (5a)$$

$$\vec{r_{FR}} = \vec{r_{C0}} - r_{C0/Fx}\hat{x}_{LDV} - r_{C0/Fy}\hat{y}_{LDV} + r_{C0/Fz}\hat{z}_{LDV} \qquad (5b)$$

where $r_{C0/Fx}$, $r_{C0/Fy}$, and $r_{C0/Fz}$ are the three components of the transducer locations in a coordinate system fixed in the SLDV. The $\hat{z}_{LDV}$ direction is the instrument's primary sensitive direction and follows the line of the returning laser beam in its zero position. The orthogonal directions are also related to beam orientation: $\hat{y}_{LDV} = \hat{b}_2$ (always) and $\hat{x}_{LDV} = \hat{b}_3$ when the 'y-scan' mirror is in its zero position.

For a single compensating measurement at the rear of the second scanning mirror (shown as 'AccR' in Figure 1), the position vector, $\vec{r_R}$, is:

$$\vec{r_R} = \vec{r_{C0}} + r_{C0/Rz}\hat{z}_{LDV} \qquad (5c)$$

where $r_{C0/Rz}$ is the z-component of the transducer location in the SLDV coordinate system. The other components of the position vector are both zero in this case. The vibration velocities, $\vec{V_{FL}}$, $\vec{V_{FR}}$ and $\vec{V_R}$, at these locations can be written as:

$$\vec{V_{FL}} = \vec{V_0} + \left(\vec{r_{C0}} + r_{C0/Fx}\hat{x}_{LDV} + r_{C0/Fy}\hat{y}_{LDV} + r_{C0/Fz}\hat{z}_{LDV}\right) \times \vec{\omega} \qquad (6a)$$

$$\vec{V_{FR}} = \vec{V_0} + \left(\vec{r_{C0}} - r_{C0/Fx}\hat{x}_{LDV} - r_{C0/Fy}\hat{y}_{LDV} + r_{C0/Fz}\hat{z}_{LDV}\right) \times \vec{\omega} \qquad (6b)$$

$$\vec{V_R} = \vec{V_0} + \left(\vec{r_{C0}} + r_{C0/Rz}\hat{z}_{LDV}\right) \times \vec{\omega} \qquad (6c)$$

With the front face accelerometer pair oriented to measure vibration in the $\hat{z}_{LDV}$ direction, the measured velocities, $U_{FL}$ and $U_{FR}$, can be written:

$$U_{FL} = \left(\hat{z}_{LDV} \cdot \vec{V_{FL}}\right) = \hat{z}_{LDV} \cdot \left(\vec{V_0} + \left(\vec{r_{C0}} + r_{C0/Fx}\hat{x}_{LDV} + r_{C0/Fy}\hat{y}_{LDV} + r_{C0/Fz}\hat{z}_{LDV}\right) \times \vec{\omega}\right) \qquad (7a)$$

$$U_{FR} = \left(\hat{z}_{LDV} \cdot \vec{V_{FR}}\right) = \hat{z}_{LDV} \cdot \left(\vec{V_0} + \left(\vec{r_{C0}} - r_{C0/Fx}\hat{x}_{LDV} - r_{C0/Fy}\hat{y}_{LDV} + r_{C0/Fz}\hat{z}_{LDV}\right) \times \vec{\omega}\right) \qquad (7b)$$

Taking the mean of these measured velocities, as required, and making use of equation (4) to obtain the correction velocity $U_F$:

$$U_F = \tfrac{1}{2}(U_{FL} + U_{FR}) = (\hat{z}_{LDV} \cdot \overrightarrow{V_{C0}}) + \tfrac{1}{2}\hat{z}_{LDV} \cdot \left((r_{C0/Fx}\hat{x}_{LDV} + r_{C0/Fy}\hat{y}_{LDV} + r_{C0/Fz}\hat{z}_{LDV}) \times \vec{\omega}\right) +$$
$$\tfrac{1}{2}\hat{z}_{LDV} \cdot \left((-r_{C0/Fx}\hat{x}_{LDV} - r_{C0/Fy}\hat{y}_{LDV} + r_{C0/Fz}\hat{z}_{LDV}) \times \vec{\omega}\right) \quad (8a)$$

Re-ordering the triple scalar products shows more readily that together they evaluate to zero because the $\hat{x}_{LDV}$ and $\hat{y}_{LDV}$ components sum to zero leaving only a cross-product between $\hat{z}_{LDV}$ and itself:

$$\vec{\omega} \cdot \left(\hat{z}_{LDV} \times (r_{C0/Fx}\hat{x}_{LDV} + r_{C0/Fy}\hat{y}_{LDV} + r_{C0/Fz}\hat{z}_{LDV}) + \hat{z}_{LDV} \times (-r_{C0/Fx}\hat{x}_{LDV} - r_{C0/Fy}\hat{y}_{LDV} + r_{C0/Fz}\hat{z}_{LDV})\right) = 0 \quad (8b)$$

Consequently:

$$U_F = (\hat{z}_{LDV} \cdot \overrightarrow{V_{C0}}) \quad (8c)$$

which is equal to the required correction of $(\hat{b}_4 \cdot \overrightarrow{V_C})$ when the laser beam is in its zero position.

Following the equivalent analysis for the rear accelerometer, the alternative correction velocity, $U_R$, can be written:

$$U_R = (\hat{z}_{LDV} \cdot \overrightarrow{V_R}) = \hat{z}_{LDV} \cdot (\overrightarrow{V_0} + (\overrightarrow{r_{C0}} + r_{C0/Rz}\hat{z}_{LDV}) \times \vec{\omega}) \quad (9a)$$

With the equivalent rearrangement, the measured velocity for this option is also shown to be equal to the required correction of $(\hat{b}_4 \cdot \overrightarrow{V_C})$ when the laser beam is in its zero position:

$$U_R = (\hat{z}_{LDV} \cdot \overrightarrow{V_{C0}}) + \hat{z}_{LDV} \cdot \left((r_{C0/Rz}\hat{z}_{LDV}) \times \vec{\omega}\right) = (\hat{z}_{LDV} \cdot \overrightarrow{V_{C0}}) \quad (9b)$$

Either correction velocity, $U_F$ or $U_R$, is then subtracted from the original SLDV measurement, $U_m$. Correction based on either option will be explored experimentally in the following section.

## 3    Experimental confirmation of measurement correction schemes

### 3.1    Experimental arrangement

The entire SLDV unit was mounted on a linear bearing, enabling vibration in the *z* direction (only). This vibration is generated via an electrodynamic shaker connected to the SLDV body by a stiff pushrod, as shown in Figure 2c. With the shaker inactive, the SLDV is (nominally) stationary. The target was another electrodynamic shaker, orientated in line with the sensor head optical axis and instrumented with an Endevco 770F-010-U-120 variable capacitance 'DC response' accelerometer, 'AccTar', as shown in Figure 2d. Again, the target is (nominally) stationary when the shaker is not active. The LDV stand-off distance (505 mm) was set to coincide with one of the recommended working distances for the NLV-2500-5 [9], which takes into account the laser beam path within the scanning head as well as the distance to the target.

The shakers are driven independently, via appropriate power amplifiers (not shown), with signals generated within the 8 input/4 output channel configuration B&K LAN-XI/Pulse LabShop data acquisition system (also not shown).

*3.2    Accelerometer sensitivity relative calibration and time delay compensation*

Following the procedure established in previous studies [2]-[4] and for optimal measurement compensation, it is good practice to confirm the accelerometer sensitivities with respect to that of the LDV but it is essential to measure the finite time delays between the accelerometer channels and that of the LDV. As shown in Figure 3a, this was practically achieved by mounting the entire accelerometer set directly onto an electrodynamic shaker. In this image, five similar Endevco transducers are shown, though only four were subsequently used. The interface between each sensor is a thin layer of synthetic beeswax, stiff enough to ensure identical vibration occurs in all units across the frequency range used. The LDV, used as the reference vibration measurement, is aligned with the accelerometers' sensitive axes with the probe laser beam focused on the top accelerometer in the stack.

[INSERT: Figure 3. Amplitude calibration check and time delay calculation; a) experimental set-up with laser beam path super-imposed, b) mean amplitude comparison with LDV signal after accelerometer *sensitivity* adjustment and c) phase difference between LDV and example accelerometer before and after *time delay* adjustment.]

A broadband random (white) noise signal to 200 Hz was generated to yield an accelerometer stack vibration level of between 1 and 10 mm/s. This level is deliberately chosen to be consistent with the level used in a common accelerometer calibrator, e.g. the Brüel & Kjaer Type 4294 [10] (albeit at single frequency i.e. 10 mm/s RMS at 159.2 Hz) and with the level that corresponds to a "slightly rough" running machine [11]. It is therefore reasonable to use in these experiments but the exact level achieved is not critical since the LDV provides the reference sensitivity against which the relative accelerometer sensitivities are to be determined.

Five sequential complex frequency spectra (200 Hz range, 0.5 Hz resolution, no overlap) were directly collected for all channels simultaneously. Frequency domain post-processing conveniently enables integration of the accelerometer signals to velocity for direct comparison with the LDV. RMS values for the frequency range between 2.5 and 100 Hz were calculated for each measurement channel and the ratio for each accelerometer with respect to the LDV was used to re-calculate the accelerometer sensitivities. The resulting small adjustments, on the order of 1.5-2%, are summarised in Table 1. Figure 3b shows a comparison of velocity spectra *after* adjustment; the target accelerometer ('AccTar') is arbitrarily selected to illustrate this. As expected, the agreement between the example accelerometer and LDV signals is very strong, indifferentiable to the eye across the entire selected frequency range.

[INSERT: Table 1. Summary of accelerometer sensitivities and inter-channel time delays with respect to the LDV.]

The same data are used to calculate the inter-channel time delays. Figure 3c shows the mean phase differences for the example accelerometer with respect to the LDV both before (dashed curve) and after (solid curve) time delay compensation. Prior to time delay compensation, there is a small positive gradient, indicating a finite time delay between the accelerometer and the LDV channel. After compensation the gradient is eliminated, indicating that the signals have been time-aligned. Table 1 also presents the time delays for the four accelerometer channels relative to the LDV channel. These relative delays come from the signal conditioning systems, not the data acquisition system, and all very similar at around 0.14 ms, which is to be expected as the accelerometers are the same model and use the same signal conditioning.

*3.3   Experimental confirmation of SLDV sensitivity to sensor head vibration and measurement correction*

Confirmation of the SLDV sensitivity to its own vibration is a logical starting point for the experimental part of this study. The SLDV shaker was driven with broadband random noise to 200 Hz, to generate vibration at around 1 mm/s RMS (level equivalent to that for a "good" running machine [11]). The target shaker was not active so the target accelerometer only picks up ambient vibration. The SLDV laser beam was incident on the 'stationary' target. Five complex frequency spectra (200 Hz range, 0.5 Hz resolution, no overlap) were captured for subsequent frequency domain processing. Figure 4a shows the average amplitude spectra for the SLDV and target accelerometer measurements. The difference is clear; the true target vibration is genuinely very low but the SLDV measurement is at a much higher level because of the vibration of the instrument itself. There are various features in each spectrum, associated with force and response in the normal way, but the shapes of the spectra are not important here. The important point is the sensitivity in the SLDV measurement to vibration of the SLDV itself and that the levels are such that compensation for this vibration is essential to yield a reliable measurement of the target vibration.

[INSERT: Figure 4. Comparison between SLDV measurement (*during vibration)* and the 'true' target vibration for a (nominally) stationary target for both correction options; a) measured averaged spectra, b) proposed correction measurements, c) corrected averaged spectra and d) averaged dB reduction plots.]

Formulating correction velocities in the manner of equation (8c) using the accelerometer pair on the front face of the SLDV body (dot-dash curve with *triangle* markers) and in the manner of equation (9b) using the single accelerometer mounted behind the final scanning mirror (dot-dash curve with *square* markers) validates the use of these equations for compensation as well as the hypothesis that sits behind them. Figure 4b shows excellent agreement between the measurements for both correction options and the original SLDV measurement. This confirms the SLDV sensitivity to its own vibration, according to equation (3c), and is the second important finding in this paper.

Notably, however, the agreement is less good between the original SLDV measurement and the correction derived from the accelerometer pair on the front face of the SLDV body in the frequency range above approximately 70 Hz. This can be attributed to the dynamic response of the relatively flexible SLDV body front panel to which the accelerometer pair is fixed. (Indeed, the agreement was improved by subsequently tightening the screws on the SLDV body i.e. by stiffening the panel). The requirement for rigid body vibration of the instrument, as explained alongside the presentation of equations (2a-c), is, to be precise, that the laser, scanning mirrors and compensation measurement location(s) move as a rigid body and the location of the single accelerometer, which is mounted to a relatively stiff mirror galvanometer mounting bracket, better fulfils this requirement. This observation has an important implication for the final choice of correction measurement location.

Notwithstanding this issue, both correction options offer significant improvement to the original problematic SLDV measurement, as shown in Figure 4c. The correction is of course less good at the higher frequencies for the accelerometer pair option but, for the single accelerometer mounted on the stiff bracket behind the final scanning mirror, correction is excellent across the entire frequency range. An exact match in the figure between the 'true' target vibration (from the target accelerometer) and the corrected SLDV measurement cannot reasonably be expected because the data are essentially a comparison between two extremely low vibration levels presented on a log scale. (Note that, while the data presented in Figure 4a-c are averaged over the five individual spectra, the correction is applied at the individual spectrum level.)

Expressing the improvement as a dB error reduction, based on equation (A5) from Appendix A, is a standard approach in such circumstances and Figure 4d shows this dB reduction for both the accelerometer pair (16.8 dB mean) and single accelerometer (26.6 dB mean) options. Such significant reductions further validate equation (3c) and confirm the effectiveness of equations (8c) and (9b) in providing the necessary correction velocity.

*3.4    Measurement correction for simultaneous SLDV and target vibration*

Having demonstrated SLDV measurement sensitivity to vibration of the instrument itself and the potential for effective correction, the next step is to examine the case of simultaneous target and instrument vibration. Independent, broadband random target and SLDV vibrations were arranged with levels in the range 1-2 mm/s RMS. As can be seen in Figure 5a, the differences between the SLDV measurement and the 'true' target vibration are evident, particularly in the frequency range up to 20

Hz. Again, there is no special significance to the spectral shape of either measurement. No control has been exerted over the particular combination of forces and responses as the correction can (and must) be applied in any circumstances.

[INSERT: Figure 5. For simultaneous target *and* SLDV sensor head vibration, comparison between measurements from the SLDV and of the 'true' target vibration; a) measured averaged spectra, b) proposed correction measurements, c) corrected averaged spectrum (single accelerometer only), d) averaged dB reduction (single accelerometer only) and e) example phase difference spectrum (single accelerometer only).]

The formulated corrections are shown in Figure 5b for comparison with the original SLDV measurement. As before, correction has been formulated from the accelerometer pair (dot-dash with triangle marker) and the single accelerometer (dot-dash with square marker). It is clear from the comparison between the uncorrected SLDV measurement and the true target vibration that correction will be significant in the region up to 20 Hz, less significant but still important up to around 50 Hz and then less important above 50 Hz where the SLDV measurement appears dominated by genuine target vibration. The flexibility of the front panel plate is again evident in the formulated accelerometer pair correction around 80-90 Hz.

Applying either of the corrections results in an impressive improvement to the SLDV measurement. Figure 5c shows this, in the interests of brevity and clarity, for the single accelerometer option only. Visually, there is excellent agreement between the corrected SLDV measurement and the true target vibration, across the entire frequency range. The dB error reduction shown in Figure 5d gives more quantitative insight. Here, the mean reduction over the displayed frequency range is 19.7 dB (15.1 dB for the accelerometer pair correction, not shown). In the range 2.5-20 Hz, where the need for significant correction was observed, the error is reduced by 26 dB, while from 20-50 Hz, where more modest correction was needed, the error is reduced by 22 dB. Above 50 Hz, where the original SLDV measurement appeared not to need significant correction, the error reduction is small, as expected, but still around 5 dB. Figure 5e shows, in this case for one of the individual spectra sets, the phase error, based on equation (A9) from Appendix A, between a corrected SLDV measurement and the true vibration measurement. Good agreement is observed across the entire frequency range, with especially good agreement in the frequencies above approximately 20 Hz where there is reasonable vibration level for *both* signals. Phase errors are larger at low frequencies where the true vibration level is low and integration noise (in the conversion of accelerometer measurements to velocity) is significant.

## 4 Experimental validation of measurement correction during scanning

### 4.1 *Measurement correction during scanning*

Demonstrating the correction for sensor head vibration of SLDV measurements is important and has not previously been reported. However, in the configuration presented so far, with the laser beam in its zero position, the measurement has been fundamentally the same as that from a fixed beam LDV. One of the main differentiators of this study is its consideration of the effect of sensor head vibration on SLDV measurements when the laser beam is oriented away from its zero position i.e. when $\hat{b}_4 \neq \hat{z}_{LDV}$ and when $\hat{b}_4$ is variable. Furthermore, it is necessary to find a means of correction that balances simplicity with effectiveness.

With reference to Figure 1 and Figure 6, the orientation vector $\hat{b}_4$ can be expanded (see Appendix B) in terms of the mirror scan angles, $\theta_y$ (first, 'y-scan' mirror) and $\theta_x$ (second, 'x-scan' mirror), as follows:

$$\hat{b}_4 = -\cos 2\theta_y \sin 2\theta_x \, \hat{x}_{LDV} + \sin 2\theta_y \, \hat{y}_{LDV} - \cos 2\theta_y \cos 2\theta_x \, \hat{z}_{LDV} \qquad (10)$$

The required correction can then be written as:

$$(\hat{b}_4 . \vec{V_C}) = -\cos 2\theta_y \sin 2\theta_x \, (\hat{x}_{LDV} . \vec{V_C}) + \sin 2\theta_y \, (\hat{y}_{LDV} . \vec{V_C}) - \cos 2\theta_y \cos 2\theta_x \, (\hat{z}_{LDV} . \vec{V_C}) \qquad (11)$$

Equation (11) shows clearly that full correction requires measurements of $\vec{V_C}$ rather than $\vec{V_{C0}}$, of three components of $\vec{V_C}$ and of mirror scan angles. However, it is also clear that proximity dictates $\vec{V_{C0}} \approx \vec{V_C}$ unless there is some reason to expect particularly high levels of angular motion. While, especially for smaller scan angles, the dominant term in the correction will be the $\hat{z}_{LDV}$ component such that:

$$(\hat{b}_4 . \vec{V_C}) \approx -\cos 2\theta_y \cos 2\theta_x \, (\hat{z}_{LDV} . \vec{V_{C0}}) \qquad (12)$$

Equation (12) indicates that correction also requires a geometrical weighting, $\cos 2\theta_y \cos 2\theta_x$, to be applied to the correction measurement made. However, this weighting is only small. Without it, the error in the compensation can be written as:

$$e = 1 - (\hat{z}_{LDV} . \vec{V_{C0}})/(\hat{b}_4 . \vec{V_C}) = 1 - \frac{1}{\cos 2\theta_y \cos 2\theta_x} \qquad (13)$$

which evaluates to, for example, only 2.2% error for 12° beam deflection for one mirror.

[INSERT: Figure 6. (a) Top, side and front view schematic diagrams showing laser beam path and correction accelerometer locations. (b) Physical set-up showing mirror rotations, beam path (super-imposed) and target.]

Consequently, the next set of experiments explore the effectiveness of a single correction measurement of $(\hat{z}_{LDV} . \vec{V_{C0}})$ for the situation where $\hat{b}_4 \neq \hat{z}_{LDV}$ and this will be undertaken with and without the geometrical weighting indicated by equation (12). The experimental set-up drives SLDV vibration in the $\hat{z}_{LDV}$ direction only (though some inevitable angular motion is also expected) so this analysis is concentrated on the consequences of measurement of a component of $\vec{V_{C0}}$ rather than of $\vec{V_C}$ and the importance of the geometrical weighting.

Figure 6 shows positive sense *x*-scan and *y*-scan mirror angles that result in positive sense laser beam deviation in the target plane in the *x* and *y* directions. The scanning mirrors are driven by application of 0.25 V per degree optical to each of the galvanometer drive amplifiers [12]. For the experiment, fixed voltages were applied independently to the scanning mirror galvanometer amplifiers to achieve 2° increments in the optical scan angle between -4° and +12° on either *x*- or *y*-scan mirrors. Measurements across the full angular ranges were not considered necessary but the -4° and -2° scenarios were included to confirm the expected symmetry in behaviour. Once the laser beam orientation had been set, the target was repositioned to maintain an optimum stand-off distance [9] and then reoriented to ensure that its vibration direction was aligned with the laser beam direction, as illustrated by the dashed line in Figure 6a and shown photographically for an example *x*-scan in Figure 6b.

Data were collected with both target and instrument undergoing broadband vibration of equivalent levels to those applied previously. The error reductions described below confirm that the proposed correction schemes are highly effective and this is the third important finding of this study. Example amplitude spectra for the two extreme scan angles, -4° and +12°, for the *y*-scan direction case and using the single accelerometer correction option are shown in Figure 7a&b. The particular corrections shown do not include the geometrical weighting but, visually, the corrections work well even for the larger scan angle. (The corresponding scenarios for the *x*-scan direction case are similar but are not shown for the sake of brevity.) For the phase error as a function of frequency, Figure 8 shows two scenarios from the +12° *y*-scan, though again all data show similar trends. The lowest overall phase error is found for the single accelerometer correction option with the geometrical weighting, while the highest overall error is for the accelerometer pair correction option without the geometrical weighting. Though not large, the differences are visible in Figure 8 which reveals a low frequency region with larger errors associated with integration noise, a mid-frequency region where errors are low in both cases and a higher frequency region where the weaker correction option shows larger errors as a consequence of noise affecting the smaller vibration amplitudes found in this range. The effect of the SLDV body response in the range 80-90 Hz, encountered throughout this study, is also apparent in the plot for the accelerometer pair.

[INSERT: Figure 7. Comparison between measurements from the vibrating SLDV and the 'true' target vibration *during scanning* for single accelerometer correction without geometrical weighting; a) -4° and b) +12°.]

[INSERT: Figure 8. Example phase difference plot comparing correction options *during scanning*.]

*4.2   Quantitative evaluation of the correction options*

To compare the correction options quantitatively and more effectively, the mean dB error reduction was calculated according to equation (A5). Appendix A further develops a total RMS phase error calculated according to equation (A11). The dB error reductions are presented in Tables 2a&b for the *x*-scan and *y*-scan configurations respectively and the total RMS phase errors are similarly presented in Tables 3a&b. The correction options are presented in descending order of effectiveness. Each column of each table represents a single measurement (average for five data captures) with five channels of data: the SLDV, the target accelerometer and the three accelerometers in use for the two correction options.

[INSERT: Table 2a: dB error reductions in the frequency range 2.5 to 100 Hz; *x*-scan.]

[INSERT: Table 2b: dB error reductions in the frequency range 2.5 to 100 Hz; *y*-scan.]

The clearest trend in the dB error reduction data is the greater reduction associated with the single accelerometer option compared to the accelerometer pair, from at least 4.4 dB up to as much as 10.3 dB. For any particular correction option, there is no clear trend with scan angle. For the options including geometrical weighting, this might be expected and suggests that the variation observed in error reduction is affected mainly by sources such as measurement noise or transducer alignments. For the options not including geometrical weighting, the effect of geometry might be expected to be apparent at greater scan angles but this trend is not apparent when observing only an individual row of the tables. However, the trend is reliably apparent (though small) in the comparison between equivalent correction options at higher scan angles. For example, in Table 2a for the single accelerometer configuration and the *x*-scan direction, inclusion of the geometrical weighting adds 0.1 dB, 0.2 dB and 0.4 dB, for scan angles of 8°, 10° and 12° respectively, to the corresponding error reductions achieved without the geometrical weighting. While for the accelerometer pair, the geometrical correction adds 0.1 dB, 0.1 dB, 0.2 dB and 0.3 dB, for scan angles of 6°, 8°, 10° and 12° respectively, to the error reductions achieved without the geometrical weighting. Similar improvement is evident in Table 2b for the *y*-scan direction.

Equation (A7b) in Appendix A allows investigation of why the effect of the geometrical correction is so small. When the geometrical correction is applied, $\alpha = 1$ and the error reduction, $R_1$, is limited just by additive noise in the measurements:

$$R_1 = -10 \log_{10}\left[\frac{\overline{n^2(t)}}{\overline{c^2(t)}}\right] \tag{14a}$$

From Tables 2a&b, an indicative calculation can be made with $R_1$ as 20 dB, i.e. $\left[\overline{\frac{n^2(t)}{c^2(t)}}\right] = 0.01$. Without the geometrical correction, $\alpha = 1/\cos 2\theta_y \cos 2\theta_x$ and the error reduction, $R_2$, can then written as:

$$R_2 = -10 \log_{10}[(1-\alpha)^2 + 0.01] \qquad (14b)$$

For a +12° $x$- or $y$-scan, $R_2$ evaluates to 19.8 dB, i.e. the effect of including the geometrical weighting is to improve the error reduction by 0.2 dB from 19.8 dB to 20 dB, consistent with the values obtained experimentally. These data and considerations confirm that the geometrical correction is working as expected and should be applied but its effect is small relative to that of routine measurement errors.

[INSERT: Table 3a: Total RMS phase errors (mrad) in the frequency range 2.5 to 100 Hz; $x$-scan.]

[INSERT: Table 3b: Total RMS phase errors (mrad) in the frequency range 2.5 to 100 Hz; $y$-scan.]

Equivalent trends can be seen in the total RMS phase error data shown in Tables 3a&b. The smaller error associated with the single accelerometer correction option relative to the accelerometer pair option is also evident in the phase error, with the single accelerometer option offering lower errors by 20 to 100 mrad. For any particular correction option, there is again no clear trend with scan angle but the benefit of inclusion of the geometrical weighting is apparent in the comparison between equivalent correction options at higher scan angles. For example, in Table 3a for the single accelerometer and the $x$-scan, including the geometrical weighing takes 10, 31 and 65 mrad, for scan angles of 8°, 10° and 12° respectively, off the RMS error found without the geometrical weighting. While for the accelerometer pair, including the geometrical weighting takes 9, 26 and 51 mrad, for scan angles of 8°, 10° and 12° respectively, off the RMS error incurred without the geometrical weighting. Similar improvement is evident in Table 3b for the $y$-scan. The data further confirm that the geometrical weighting should be applied. Though small, its measurable benefit is provided at minimal additional cost. Demonstrating that the single accelerometer correction with geometrical weighting provides the best outcome is a further important finding of this study.

*4.3   Enhanced three component correction*

While it is clear that the dominant correction component will always be that associated with $\hat{z}_{LDV}$-vibration because $\hat{b}_4$ is always closest to $\hat{z}_{LDV}$, equation (11) indicates that, for vibration components of similar magnitude, the $x$-correction will be $\tan 2\theta_x$ of the $z$-correction and the $y$-correction will be $\tan 2\theta_y / \cos 2\theta_x$ of the $z$-correction. For a 12° beam deflection, these are both slightly in excess of 20% so, while the need for an enhanced correction is very dependent on the particular measurement scenario, these values do suggest that there is merit for a robust correction setup.

The approach is not, as might immediately be expected, to use tri-axial transducers for compensation since a single sensor location (remote from C itself) cannot satisfy the requirements of all three correction components. For example, an *x*-measurement at the AccR location would give:

$$(\hat{x}_{LDV} \cdot \vec{V_R}) = \hat{x}_{LDV} \cdot (\vec{V_0} + (\vec{r_{C0}} + r_{C0/Rz}\hat{z}_{LDV}) \times \vec{\omega}) = (\hat{x}_{LDV} \cdot \vec{V_{C0}}) + \vec{\omega} \cdot (\hat{x}_{LDV} \times r_{C0/Rz}\hat{z}_{LDV}) \quad (15)$$

in which a residual sensitivity to any angular velocity is clear. The correct location, identified in Figure 6a (Front view) as 'AccRx', is a different one, with only a $\hat{x}_{LDV}$ component in its position relative to C0. Writing this as $r_{C0/Rx}\hat{x}_{LDV}$, this additional single-axis measurement, $U_{Rx}$, can be written as:

$$U_{Rx} = (\hat{x}_{LDV} \cdot \vec{V_{Rx}}) = \hat{x}_{LDV} \cdot (\vec{V_0} + (\vec{r_{C0}} + r_{C0/Rx}\hat{x}_{LDV}) \times \vec{\omega})$$

$$= (\hat{x}_{LDV} \cdot \vec{V_{C0}}) + \hat{x}_{LDV} \cdot (r_{C0/Rx}\hat{x}_{LDV} \times \vec{\omega}) = (\hat{x}_{LDV} \cdot \vec{V_{C0}}) \quad (16a)$$

The equivalent *y*-correction, $U_{Ry}$, requires a third single-axis measurement at a location, 'AccRy' (see Figure 6a (Side view)), with a position vector $r_{C0/Ry}\hat{y}_{LDV}$ relative to C0:

$$U_{Ry} = \hat{y}_{LDV} \cdot (\vec{V_0} + (\vec{r_{C0}} + r_{C0/Ry}\hat{y}_{LDV}) \times \vec{\omega})$$

$$= (\hat{y}_{LDV} \cdot \vec{V_{C0}}) + \hat{y}_{LDV} \cdot (r_{C0/Ry}\hat{y}_{LDV} \times \vec{\omega}) = (\hat{y}_{LDV} \cdot \vec{V_{C0}}) \quad (16b)$$

Note also that, in Figure 6a (Side view), the location previously identified as AccR becomes 'AccRz' to emphasise that correction at this location is only effective for *z*-vibration and to be consistent with the labelling adopted for this enhanced correction. The incident points, C0 and C, are also shown on this view.

Substituting equations (9b) and (16a&b) into equation (11) leads to an improved compensation as follows:

$$(\hat{b}_4 \cdot \vec{V_C}) \approx -\cos 2\theta_y \sin 2\theta_x \, U_{Rx} + \sin 2\theta_y \, U_{Ry} - \cos 2\theta_y \cos 2\theta_x \, U_{Rz} \quad (17)$$

The practical implementation of this correction will be the subject of further work.

*4.4   Error associated with fixed placement of correction accelerometers*

The standard and enhanced corrections proposed by equations (11) and (17) still rely on $\vec{V_C} \approx \vec{V_{C0}}$ so the final part of this paper will estimate the typical error associated with this approximation. The derivation of the position vector for the exact incident point is set out in Appendix C, which shows:

$$\vec{r_C} = \vec{r_{C0}} + |\overrightarrow{BC0}| \tan 2\theta_y \, \hat{y}_{LDV} \tag{18}$$

The error associated with measurement at C0 rather than C can be written as:

$$e = \frac{\hat{b}_4.\vec{V_C} - \hat{b}_4.\vec{V_{C0}}}{\hat{b}_4.\vec{V_C}} \tag{19a}$$

which can then be re-written, by substituting equations (2c), (4) and (18) into equation (19a), as:

$$e = \frac{|\overrightarrow{BC0}| \tan 2\theta_y \vec{\omega}.(\hat{b}_4 \times \hat{y}_{LDV})}{\hat{b}_4.\vec{V_{C0}} + |\overrightarrow{BC0}| \tan 2\theta_y \vec{\omega}.(\hat{b}_4 \times \hat{y}_{LDV})} \tag{19b}$$

The terms within equation (19b) can be expanded, based on equation (10):

$$(\hat{b}_4.\vec{V_{C0}}) = -\cos 2\theta_y \sin 2\theta_x (\hat{x}_{LDV}.\vec{V_{C0}}) + \sin 2\theta_y (\hat{y}_{LDV}.\vec{V_{C0}}) - \cos 2\theta_y \cos 2\theta_x (\hat{z}_{LDV}.\vec{V_{C0}}) \tag{20a}$$

$$(\hat{b}_4 \times \hat{y}_{LDV}) = -\cos 2\theta_y \sin 2\theta_x \, \hat{z}_{LDV} + \cos 2\theta_y \cos 2\theta_x \, \hat{x}_{LDV} \tag{20b}$$

The angular vibration velocity can also be expanded into components, $\omega_x$, $\omega_y$ and $\omega_z$, as follows:

$$\vec{\omega} = (\omega_x \hat{x}_{LDV} + \omega_y \hat{y}_{LDV} + \omega_z \hat{z}_{LDV}) \tag{20c}$$

such that:

$$|\overrightarrow{BC0}| \tan 2\theta_y \, \vec{\omega}.(\hat{b}_4 \times \hat{y}_{LDV}) = |\overrightarrow{BC0}| \tan 2\theta_y (\cos 2\theta_y \cos 2\theta_x \, \omega_x - \cos 2\theta_y \sin 2\theta_x \, \omega_z) \tag{20d}$$

Substituting equations (20a) and (20d) into equation (19b) allows a full analysis of the likely error for any given scenario but a typical error can be estimated through small angle approximations, with additional recognition that $(\hat{z}_{LDV}.\vec{V_{C0}}) \gg |\overrightarrow{BC0}| 2\theta_y \omega_x$ for typical values, to yield:

$$e \approx \frac{|\overrightarrow{BC0}| 2\theta_y \omega_x}{-(\hat{z}_{LDV}.\vec{V_{C0}})} \tag{21}$$

Equation (19b) shows that if the SLDV angular vibration velocity is zero then the associated measurement error is zero and this is, of course, confirmed in the approximate expression of equation (21). Similarly, the importance of minimising $|\overrightarrow{BC0}|$ to minimise the correction error is clear in equation (19b) and confirmed in equation (21). In this experimental set-up, i.e. with $|\overrightarrow{BC0}|$ of 33.81 mm [8] and a scan angle of 12° (optical), vibration amplitudes of 1 mm/s for the $\hat{z}_{LDV}$ component of $\vec{V_{C0}}$ and of 1 mrad/s for $\omega_x$ were apparent. If these vibrations were in phase (for the sake of this indicative calculation), the measurement error associated with measurement of $(\hat{z}_{LDV}.\vec{V_{C0}})$ rather than of $(\hat{z}_{LDV}.\vec{V_C})$ is approximately 0.7%. This would place an upper limit on the error reduction of 43 dB so it is quite possible that this is another (probably secondary) source of error contributing to the experimental error reductions shown in Tables 2a&b, alongside the measurement noise discussed earlier in the paper.

# 5    Conclusions

When the optical head of a laser Doppler vibrometer (LDV) is itself subject to vibration, the measurements made are sensitive to this motion in the same way that they are sensitive to target motion due to the relative, rather than absolute, nature of LDV measurement. Both mathematically and experimentally, this paper has provided the first rigorous confirmation of this effect for the Scanning LDV. Furthermore, this paper has shown that the sensitivity to vibration of the instrument itself can be effectively compensated by additional measurement(s) of the instrument motion. In particular, it has been shown that, for arbitrary vibration and any scanning head configuration, this compensation does not require individual measurements on the LDV body and each beam steering mirror but that it can be achieved solely by measurements associated with the incident point on the final beam steering mirror and this has significant practical implications.

Two readily deployable schemes using DC-response accelerometers were considered: one based on an accelerometer pair mounted to the SLDV body front panel and a second with a single accelerometer mounted directly behind the final ($x$-scan) beam steering mirror. The sensitive axes of these accelerometers were aligned with the outgoing laser beam in its zero (unscanned) orientation. In either case, it is essential that compensation is made for inter-channel time delays. Accelerometers were chosen for their convenience and reliability but, while this study has been indisputably successful, the usual challenges from integration noise at low frequencies have been the primary factor limiting the error reductions achieved in practice.

In experiments with a vibrating SLDV but a stationary target, error reduction (2.5-100 Hz range) was 17 dB (accelerometer pair) and 27 dB (single accelerometer), with the laser beam in its zero position. In a scenario more relevant to the real-world, where both target and SLDV vibrate, error reduction (2.5-100 Hz range) was consistently around 14 dB (accelerometer pair) and 20 dB (single accelerometer) across beam scan angles from -4° to 12°. RMS phase error was also calculated, showing lower errors for the single accelerometer correction by up to 100 mrad across the same range of scan angles. These observations make clear that the single accelerometer correction is favoured because it offers lower measurement noise (because one transducer is used rather than two) and is a simpler and cheaper way forward. The results also emphasise the importance of the location at which to mount the correction transducer(s) such that LDV body, mirrors and correction measurement location(s) vibrate together as a rigid body.

Though the beam was scanned in these measurements, the correction measurements were made with single-axis accelerometers with their sensitive axes in fixed orientations. Developing the correction approach for scenarios where the laser beam is scanned is a major and unique contribution of this article. Two particular issues were investigated.

The first was that the mathematical derivation of the required correction indicated the need for a small geometrical weighting and so its importance was explored. The experiments found that, at scan angles from 6° upwards, the benefit of using the geometrical correction became apparent though, with error reductions of just a few tenths of a dB and several tens of mrad, the benefit appears small in practice. Nonetheless, it can be implemented in post-processing for zero cost and its use is recommended.

The second is that a full correction for arbitrary instrument vibration requires two additional correction measurement locations, one for each of *x* and *y* directions, to complement the primary correction for the *z*-vibration. The need for this enhanced correction was demonstrated theoretically and will be explored experimentally in further work.

In either the simple or enhanced approaches, it was recognised that all of the correction measurement locations relied on an assumption that the incident point of the laser beam on the final (*x*-scan) mirror did not change significantly during scanning. In the experiments reported here, the position (C) moved by 7 mm from the zero position (C0) for the largest scan angle used (12°). An associated error of 0.7% was estimated and shown to be a secondary contributor to the remaining error after correction. The error is directly proportional to the distance between the two scanning mirror axes and so it is recommended that this is minimised in future optical configurations. Beyond this, the complexity associated with trying to measure at point C is not felt to be justified by the potential additional error reduction and, on the basis of this study, measurement at CO is regarded as acceptable.

In summary, correction of SLDV measurements affected by sensor head vibration has been shown to be essential and achievable, and schemes have been proposed that are suitable for retro-fitting to existing instruments or inclusion in future instrument designs.

**Appendices**

*Appendix A: Definition of amplitude and phase errors*

Measurement errors can be quantified as the average over time of the square of the differences between the measurement under scrutiny and a 'true' measurement. For this work, this error can be written in terms of a corrected measurement, $U_{corr}$, and a true measurement $U_{true}$, each a function of time, as $\overline{(U_{corr}(t) - U_{true}(t))^2}$. It is convenient to quantify this error as a proportion of the error associated with the original (uncorrected) SLDV measurement, $U_m$, and to express this as a dB error reduction, $R$, as follows:

$$R = -10 \log_{10} \frac{\overline{(U_{corr}(t) - U_{true}(t))^2}}{\overline{(U_m(t) - U_{true}(t))^2}} \tag{A1}$$

in which the minus sign causes an effective reduction to take a positive dB value. The individual velocities can be considered as a sum of sine and cosine components across N individual frequencies, for example:

$$U_{corr}(t) = \sum_{n=1}^{N} A_{corr}(n) \cos n\omega t + B_{corr}(n) \sin n\omega t \tag{A2}$$

This suits the data capture in the experiments conducted here since each cosine coefficient, $A_{corr}(n)$, is the real part of the spectral component at frequency $n\omega$ and each sine coefficient, $B_{corr}(n)$, is the corresponding imaginary part. Combining velocities in the manner of equation (A1):

$$\left(U_{corr}(t) - U_{true}(t)\right)^2 =$$

$$\left(\sum_{n=1}^{N}\left(A_{corr}(n) - A_{true}(n)\right) \cos n\omega t + \left(B_{corr}(n) - B_{true}(n)\right) \sin n\omega t\right)^2 \tag{A3}$$

in which $A_{true}(n)$ and $B_{true}(n)$ are the real and imaginary parts of the $n$th spectral component of the true velocity measurement.

Expansion of the right-hand side of equation (A3) results in cross terms between all of the individual elements in the sum including the squares of each individual sin and cos term. It is only these latter terms that retain a non-zero value when the time average is taken. All of the cross terms, either between components at different frequencies or between sin and cos terms at the same frequency, average to zero over time. Consequently, the mean square error can be written as:

$$\overline{\left(U_{corr}(t) - U_{true}(t)\right)^2} = \frac{1}{2}\sum_{n=1}^{N}\left(A_{corr}(n) - A_{true}(n)\right)^2 + \left(B_{corr}(n) - B_{true}(n)\right)^2 \tag{A4}$$

Re-writing equation (A1) in the form of equation (A4) allows calculation of a dB error reduction either at an individual frequency or for any chosen frequency interval:

$$R = -10 \log_{10} \frac{\sum_{n=1}^{N}\left(A_{corr}(n) - A_{true}(n)\right)^2 + \left(B_{corr}(n) - B_{true}(n)\right)^2}{\sum_{n=1}^{N}\left(A_m(n) - A_{true}(n)\right)^2 + \left(B_m(n) - B_{true}(n)\right)^2} \tag{A5}$$

in which $A_m(n)$ and $B_m(n)$ are the real and imaginary parts of the $n$th spectral component of the original SLDV measurement.

Equation (A5) can be further developed to explore the observation that the geometrical weighting makes only a small difference, even at the higher scan angles. The correction applied, $c(t)$, can be regarded as having a proportional error, mainly related to geometry and denoted by the constant $\alpha$, and an additive error associated with measurement noise, $n(t)$.

$$c(t) = \alpha C(t) + n(t) \tag{A6a}$$

in which $C(t)$ is the perfect correction and is related to velocities as follows:

$$U_m(t) - C(t) = U_{true}(t) \tag{A6b}$$

Substituting in equations (A6a&b), equation (A1) can be re-written as:

$$R = -10\log_{10}\frac{\overline{((1-\alpha)C(t)+n(t))^2}}{\overline{C^2(t)}} \tag{A7a}$$

Expansion of the logarithm argument, noting that cross terms between $C(t)$ and $n(t)$ reduce to zero through the time average, leads to the following:

$$R = -10\log_{10}\left[(1-\alpha)^2 + \frac{\overline{n^2(t)}}{\overline{C^2(t)}}\right] \tag{A7b}$$

Phase error has to be considered spectral component by spectral component. At the *n*th frequency component, the phase error, $\Delta\varphi(n)$, can be written as follows:

$$\Delta\varphi(n) = \varphi_{corr}(n) - \varphi_{true}(n) \tag{A8}$$

in which $\varphi_{corr}(n)$ and $\varphi_{true}(n)$ are the phases of the *n*th frequency components of the corrected and true velocities respectively. Across a single spectrum it is possible to calculate an RMS phase error as follows:

$$RMS[\Delta\varphi] = \sqrt{\frac{1}{N}\sum_{n=1}^{N}\Delta\varphi^2(n)} \tag{A9}$$

An RMS error for the *n*th spectral component across multiple (*P*) runs, $RMS[\Delta\varphi(n)]_P$, can be calculated based on the phase error at that spectral component for each run $p$, $\Delta\varphi(p,n)$, as follows:

$$RMS[\Delta\varphi(n)]_P = \sqrt{\frac{1}{P}\sum_{p=1}^{P}\Delta\varphi^2(p,n)} \tag{A10}$$

and the RMS values at each spectral component can be combined in the manner of equation (A9) to give a total RMS phase error, $RMS[\Delta\varphi]_P$:

$$RMS[\Delta\varphi]_P = \sqrt{\frac{1}{N}\sum_{n=1}^{N}(RMS[\Delta\varphi(n)]_P)^2} \tag{A11}$$

The phase error analysis considers a RMS phase error calculated according to equation (A11). It is not legitimate to calculate an arithmetic mean value at each spectral component from P runs because of the circular nature of a phase calculation. This is exemplified by the case where one run results in a phase error of almost pi and a second run results in an error of almost -pi. The genuine error is approximately pi (or -pi) but the arithmetic mean would evaluate erroneously to zero.

*Appendix B: Calculation of laser beam direction unit vector, $\hat{b}_4$, during scanning*

From Figure 1, the unit vectors describing laser beam orientation can be written as:

$$\hat{b}_1 = -\hat{z}_{LDV} \tag{B1}$$

$$\hat{b}_2 = \hat{y}_{LDV} \tag{B2}$$

Following reflection at the first mirror, the new laser beam orientation, $\hat{b}_3$, can be written in terms of the mirror normal at point B, $\hat{n}_B$, as [7]:

$$\hat{b}_3 = \hat{b}_2 - 2(\hat{b}_2 \cdot \hat{n}_B)\hat{n}_B \tag{B3}$$

$\hat{n}_B$ is written as an initial orienttaion in the $-\hat{y}_{LDV}$ direction followed by an anti-clockwise rotation of $(45 + \theta_y)$ around the $\hat{z}_{LDV}$ axis. This rotation is incorporated using a rotation matrix [13]:

$$\hat{n}_B = [\hat{x}_{LDV} \quad \hat{y}_{LDV} \quad \hat{z}_{LDV}] \begin{bmatrix} \cos(45+\theta_y) & -\sin(45+\theta_y) & 0 \\ \sin(45+\theta_y) & \cos(45+\theta_y) & 0 \\ 0 & 0 & 1 \end{bmatrix} \begin{bmatrix} 0 \\ -1 \\ 0 \end{bmatrix}$$

$$= \sin(45+\theta_y)\hat{x}_{LDV} - \cos(45+\theta_y)\hat{y}_{LDV} \tag{B4}$$

Substituting equation (B4) into equation (B3) and simplifying reveals that:

$$\hat{b}_3 = \cos 2\theta_y \, \hat{x}_{LDV} + \sin 2\theta_y \, \hat{y}_{LDV} \tag{B5}$$

Using the same approach for reflection from the second scanning mirror, the final beam orientation, $\hat{b}_4$, can be written in terms of the mirror normal at incident point C, $\hat{n}_C$, as:

$$\hat{b}_4 = \hat{b}_3 - 2(\hat{b}_3 \cdot \hat{n}_C)\hat{n}_C \tag{B6}$$

$\hat{n}_C$ is written as an initial orientation in the $-\hat{z}_{LDV}$ direction followed by an anti-clockwise rotation of $(45 + \theta_x)$ around the $\hat{y}_{LDV}$ axis. This rotation is also incorporated using a rotation matrix:

$$\hat{n}_C = [\hat{x}_{LDV} \quad \hat{y}_{LDV} \quad \hat{z}_{LDV}] \begin{bmatrix} \cos(45+\theta_x) & 0 & \sin(45+\theta_x) \\ 0 & 1 & 0 \\ -\sin(45+\theta_x) & 0 & \cos(45+\theta_x) \end{bmatrix} \begin{bmatrix} 0 \\ 0 \\ -1 \end{bmatrix}$$

$$= -\sin(45+\theta_x)\hat{x}_{LDV} - \cos(45+\theta_x)\hat{z}_{LDV} \tag{B7}$$

Substituting equation (B7) into equation (B6) and simplifying reveals that:

$$\hat{b}_4 = -\cos 2\theta_Y \sin 2\theta_x \, \hat{x}_{LDV} + \sin 2\theta_Y \, \hat{y}_{LDV} - \cos 2\theta_y \cos 2\theta_x \, \hat{z}_{LDV} \tag{B8}$$

*Appendix C: Derivation of the position vector for the incident point on y-scan mirror $\vec{r_C}$, during scanning*

The position vector for the incident point on the mirror, $\vec{r_C}$, can be written in terms of a known point along the beam, $\vec{r_B}$, which is also the incident point on the previous mirror, the incoming beam orientation, $\hat{b}_3$, the position vector of a chosen reference point on the mirror, $\vec{r_{C0}}$, and the unit vector for the mirror normal at the incident point, $\hat{n}_C$, [7]:

$$\vec{r_C} = \vec{r_B} + \left[\frac{(\vec{r_{C0}} - \vec{r_B}).\hat{n}_C}{\hat{b}_3.\hat{n}_C}\right]\hat{b}_3 \tag{C1}$$

From the geometry in Figure 1, the following relationship can be written:

$$(\vec{r_{C0}} - \vec{r_B}) = |\overrightarrow{BC0}|\hat{x}_{LDV} \tag{C2}$$

Substituting equations (C2), (B5) and (B7) into equation (C1) gives:

$$\vec{r_C} = \vec{r_B} + \left[\frac{|\overrightarrow{BC0}|\hat{x}_{LDV}.(-\sin(45+\theta_x)\hat{x}_{LDV} - \cos(45+\theta_x)\hat{z}_{LDV})}{(-\sin(45+\theta_x)\cos 2\theta_y)}\right]\left(\cos 2\theta_y\,\hat{x}_{LDV} + \sin 2\theta_y\,\hat{y}_{LDV}\right) \tag{C3a}$$

which simplifies to:

$$\vec{r_C} = \vec{r_{C0}} + |\overrightarrow{BC0}|\tan 2\theta_y\,\hat{y}_{LDV} \tag{C3b}$$

## Acknowledgements

The authors wish to acknowledge the technical contributions made by Mr Chris Chapman, then Scientific Officer, Dynamics & Mechanics of Solids Laboratory and by Engineering Workshop colleagues within the Faculty of Engineering and IT, University of Technology Sydney. The authors also wish to acknowledge the contribution of Mr Abdel Darwish to manuscript integrity verification.

**Table captions**

Table 1. Summary of accelerometer sensitivities and inter-channel time delays with respect to the LDV.

Table 2a: dB error reductions in the frequency range 2.5 to 100 Hz; *x*-scan.

Table 2b: dB error reductions in the frequency range 2.5 to 100 Hz; *y*-scan.

Table 3a: Total RMS phase errors (mrad) in the frequency range 2.5 to 100 Hz; *x*-scan.

Table 3b: Total RMS phase errors (mrad) in the frequency range 2.5 to 100 Hz; *y*-scan.

**Figure captions**

Figure 1. Experimental arrangement schematic showing laser beam path and local coordinate system.

Figure 2. Experimental arrangement *physical set-up*; a) LDV body mounted to bespoke SLDV assembly, b) top view with SLDV cover removed showing 'AccR', 'AccFL' and 'AccFR', and 'AccTar' with laser beam path super-imposed, c) side view showing SLDV assembly front panel with 'AccFL' & 'AccFR' and shaker with laser beam path super-imposed and d) close-up of vibrating target with 'AccTar'.

Figure 3. Amplitude calibration check and time delay calculation; a) experimental set-up with laser beam path super-imposed, b) mean amplitude comparison with LDV signal after accelerometer *sensitivity* adjustment and c) phase difference between LDV and example accelerometer before and after *time delay* adjustment.

Figure 4. Comparison between SLDV measurement (*during vibration)* and the 'true' target vibration for a (nominally) stationary target for both correction options; a) measured averaged spectra, b) proposed correction measurements, c) corrected averaged spectra and d) averaged dB reduction plots.

Figure 5. For simultaneous target *and* SLDV sensor head vibration, comparison between measurements from the SLDV and of the 'true' target vibration; a) measured averaged spectra, b) proposed correction measurements, c) corrected averaged spectrum (single accelerometer only), d) averaged dB reduction (single accelerometer only) and e) example phase difference spectrum (single accelerometer only).

Figure 6. (a) Top, side and front view schematic diagrams showing laser beam path and correction accelerometer locations. (b) Physical set-up showing mirror rotations, beam path (super-imposed) and target.

Figure 7. Comparison between measurements from the vibrating SLDV and the 'true' target vibration *during scanning* for single accelerometer correction without geometrical weighting; a) -4° and b) +12°.

Figure 8. Example phase difference plot comparing correction options *during scanning*.

# Tables

Table 1. Summary of accelerometer sensitivities and inter-channel time delays with respect to the LDV.

| Transducer location ID | Hardware ch. | Software ch. | Acc. model | Acc. s/n | Initial sens. (mV/m/s^2) | Revised sens. (mV/m/s^2) | Time delay vs. LDV (ms) |
|---|---|---|---|---|---|---|---|
| AccFL | Ch3 | Acc3 | 770F-10-U-120 | 10013 | 21.89 | 22.23 | 0.137 |
| AccFR | Ch4 | Acc4 | 770F-10-U-120 | 10046 | 22.21 | 22.62 | 0.141 |
| AccR | Ch1 | Acc1 | 770F-10-U-120 | 10047 | 22.13 | 22.57 | 0.140 |
| AccTar | Ch2 | Acc2 | 770F-10-U-120 | 10048 | 21.29 | 21.76 | 0.140 |

Table 2a: dB error reductions in the frequency range 2.5 to 100 Hz; *x*-scan.

| Correction option | | Scan angle (deg) | | | | | | | | |
|---|---|---|---|---|---|---|---|---|---|---|
| Accelerometer(s) | Geometrical weighting | -4 | -2 | 0 | 2 | 4 | 6 | 8 | 10 | 12 |
| Single | With | 19.7 | 20.3 | 20.9 | 19.1 | 19.5 | 20.1 | 20.7 | 20.3 | 19.7 |
| Single | Without | 19.7 | 20.3 | 20.9 | 19.1 | 19.5 | 20.1 | 20.6 | 20.1 | 19.3 |
| Pair | With | 13.1 | 14.0 | 15.8 | 13.4 | 15.2 | 14.4 | 15.4 | 13.2 | 14.9 |
| Pair | Without | 13.1 | 14.0 | 15.8 | 13.4 | 15.2 | 14.3 | 15.3 | 13.0 | 14.6 |

Table 2b: dB error reductions in the frequency range 2.5 to 100 Hz; *y*-scan.

| Correction option | | Scan angle (deg) | | | | | | | | |
|---|---|---|---|---|---|---|---|---|---|---|
| Accelerometer(s) | Geometrical weighting | -4 | -2 | 0 | 2 | 4 | 6 | 8 | 10 | 12 |
| Single | With | 21.7 | 23.9 | 21.1 | 21.3 | 22.2 | 21.0 | 21.0 | 20.4 | 19.0 |
| Single | Without | 21.7 | 23.9 | 21.1 | 21.3 | 22.2 | 21.0 | 20.9 | 20.3 | 18.9 |
| Pair | With | 13.8 | 13.6 | 13.0 | 14.9 | 13.7 | 14.8 | 13.1 | 15.1 | 13.3 |
| Pair | Without | 13.8 | 13.6 | 13.0 | 14.9 | 13.6 | 14.7 | 13.0 | 14.9 | 13.1 |

Table 3a: Total RMS phase errors (mrad) in the frequency range 2.5 to 100 Hz; $x$-scan.

| Correction option | | Scan angle (deg) | | | | | | | | |
|---|---|---|---|---|---|---|---|---|---|---|
| Accelerometer(s) | Geometrical weighting | -4 | -2 | 0 | 2 | 4 | 6 | 8 | 10 | 12 |
| Single | With | 264 | 210 | 290 | 219 | 302 | 375 | 321 | 200 | 274 |
| Single | Without | 262 | 209 | 290 | 218 | 303 | 384 | 331 | 231 | 329 |
| Pair | With | 367 | 274 | 315 | 262 | 353 | 414 | 356 | 227 | 311 |
| Pair | Without | 362 | 274 | 315 | 262 | 354 | 420 | 365 | 253 | 362 |

Table 3b: Total RMS phase errors (mrad) in the frequency range 2.5 to 100 Hz; $y$-scan.

| Correction option | | Scan angle (deg) | | | | | | | | |
|---|---|---|---|---|---|---|---|---|---|---|
| Accelerometer(s) | Geometrical weighting | -4 | -2 | 0 | 2 | 4 | 6 | 8 | 10 | 12 |
| Single | With | 454 | 337 | 272 | 210 | 370 | 325 | 325 | 288 | 306 |
| Single | Without | 458 | 338 | 272 | 210 | 372 | 325 | 330 | 305 | 348 |
| Pair | With | 472 | 402 | 289 | 287 | 424 | 411 | 365 | 315 | 367 |
| Pair | Without | 476 | 402 | 289 | 287 | 424 | 410 | 373 | 334 | 402 |

**Figures**

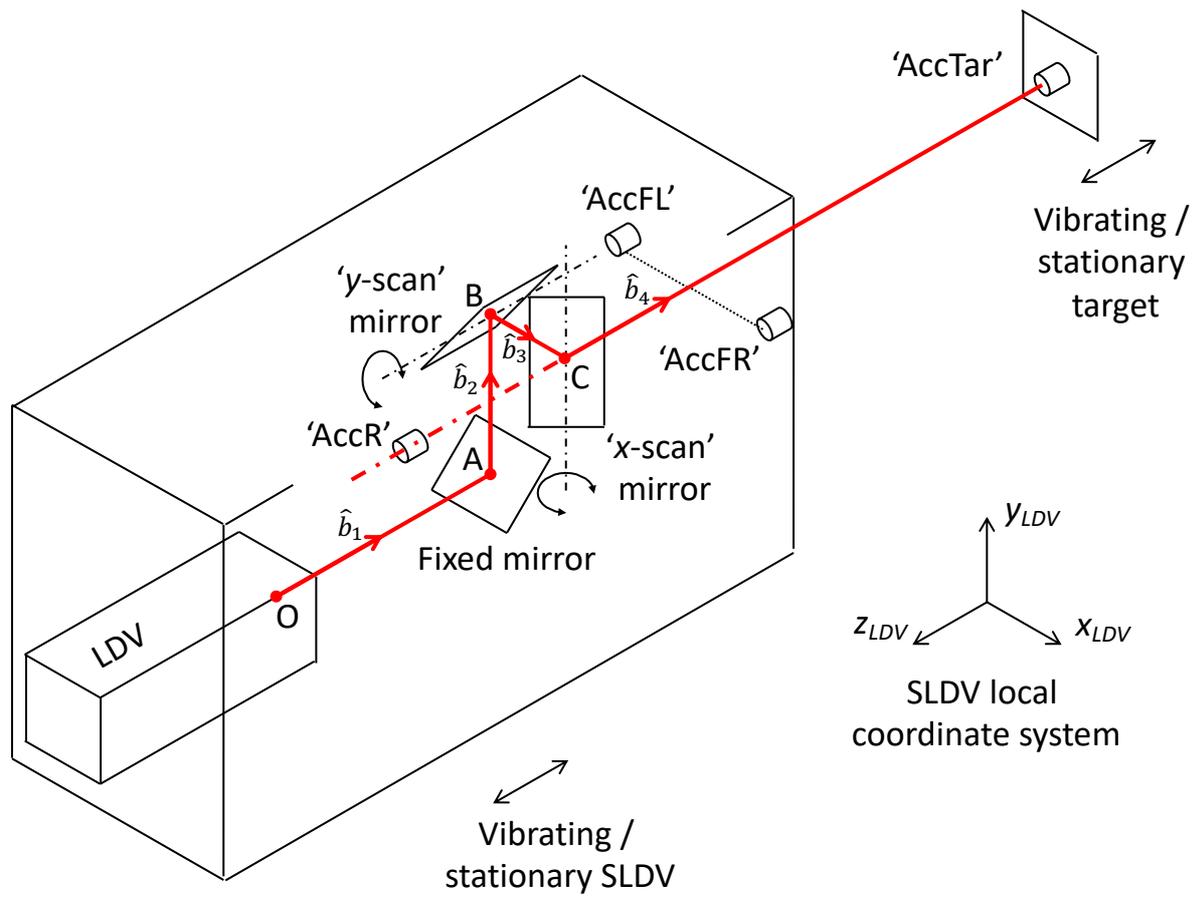

Figure 1. Experimental arrangement schematic showing laser beam path and local coordinate system.

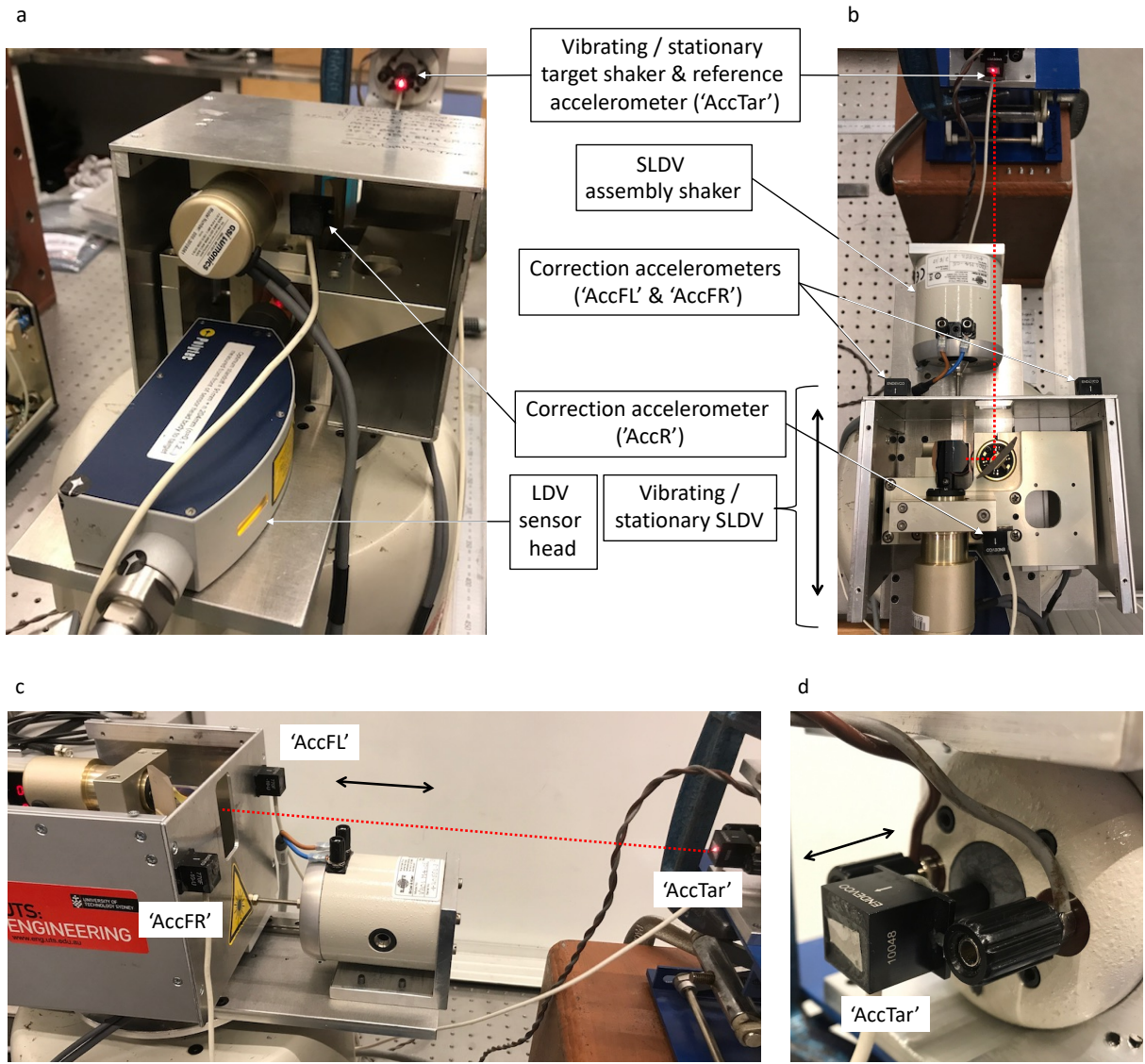

Figure 2. Experimental arrangement *physical set-up*; a) LDV body mounted to bespoke SLDV assembly, b) top view with SLDV cover removed showing 'AccR', 'AccFL' and 'AccFR', and 'AccTar' with laser beam path super-imposed, c) side view showing SLDV assembly front panel with 'AccFL' & 'AccFR' and shaker with laser beam path super-imposed and d) close-up of vibrating target with 'AccTar'.

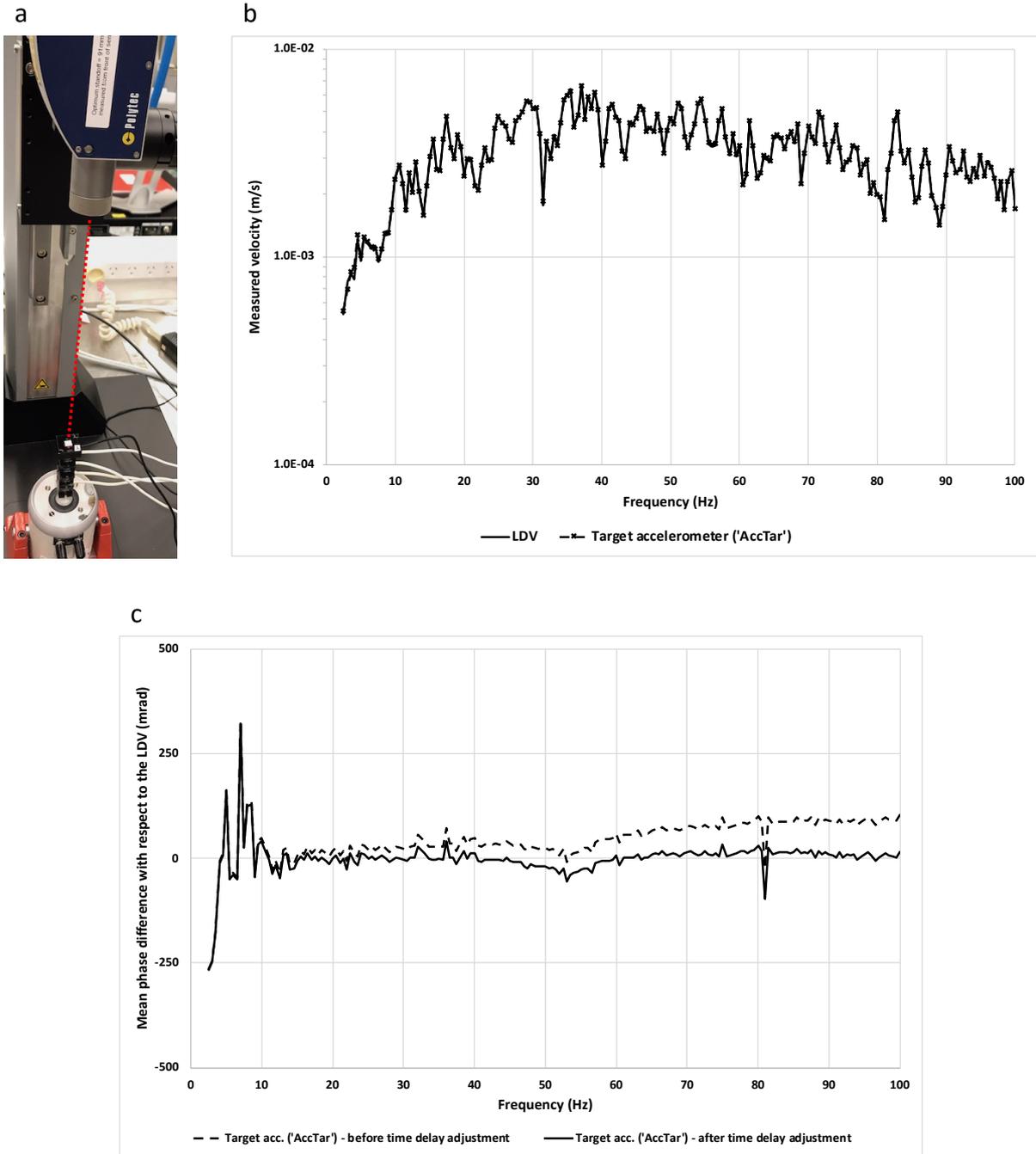

Figure 3. Amplitude calibration check and time delay calculation; a) experimental set-up with laser beam path super-imposed, b) mean amplitude comparison with LDV signal after accelerometer *sensitivity* adjustment and c) phase difference between LDV and example accelerometer before and after *time delay* adjustment.

a

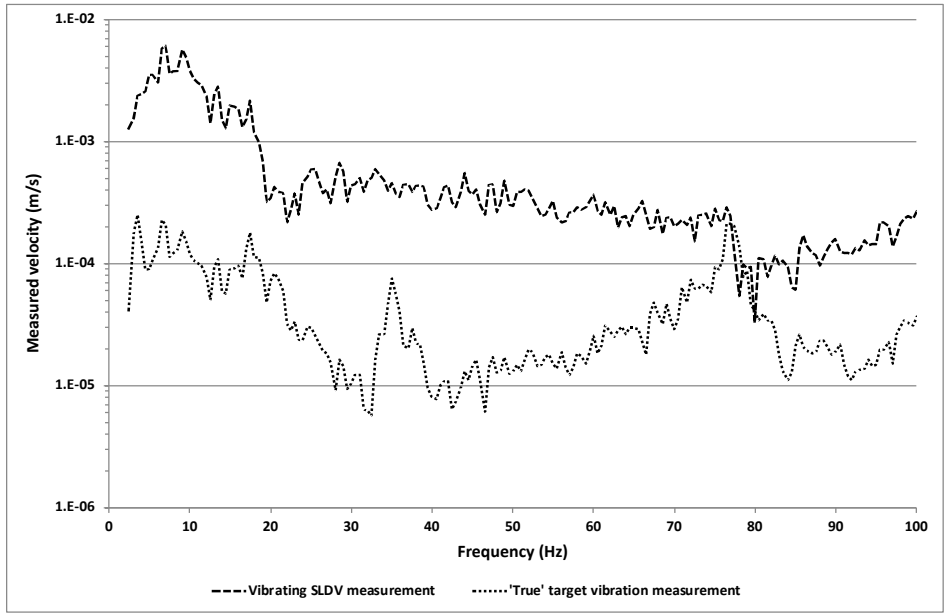

b

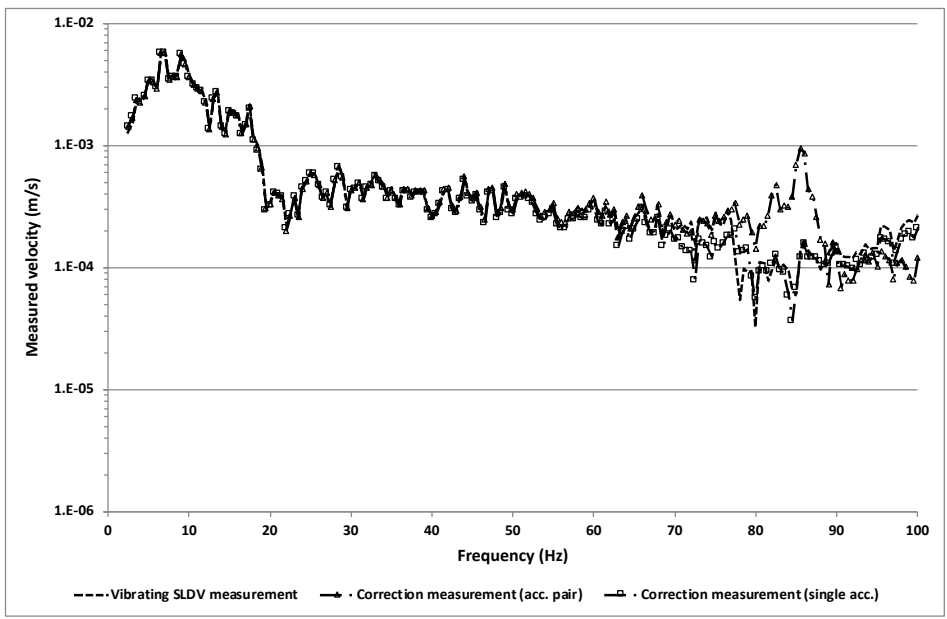

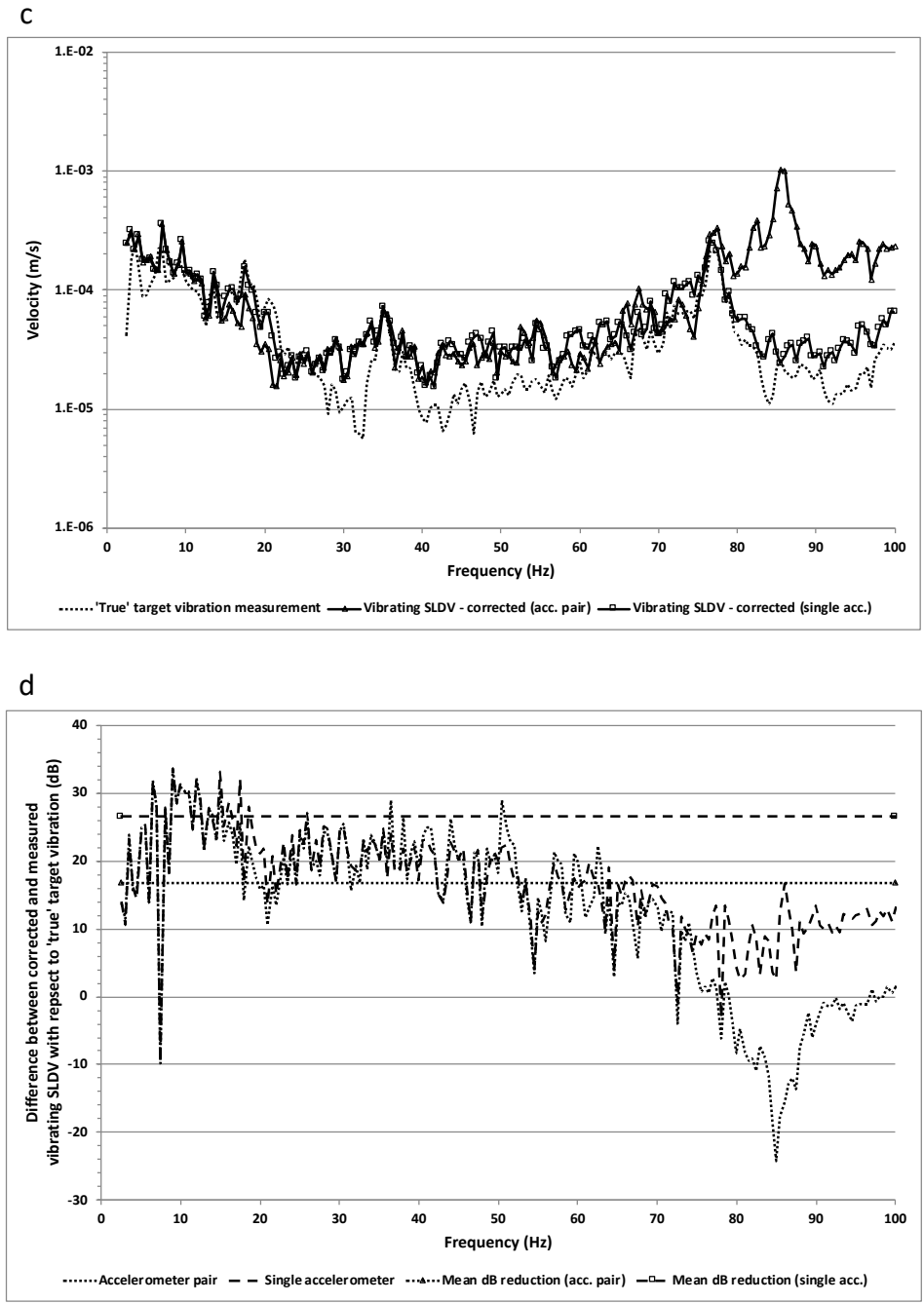

Figure 4. Comparison between SLDV measurement (*during vibration)* and the 'true' target vibration for a (nominally) stationary target for both correction options; a) measured averaged spectra, b) proposed correction measurements, c) corrected averaged spectra and d) averaged dB reduction plots.

a

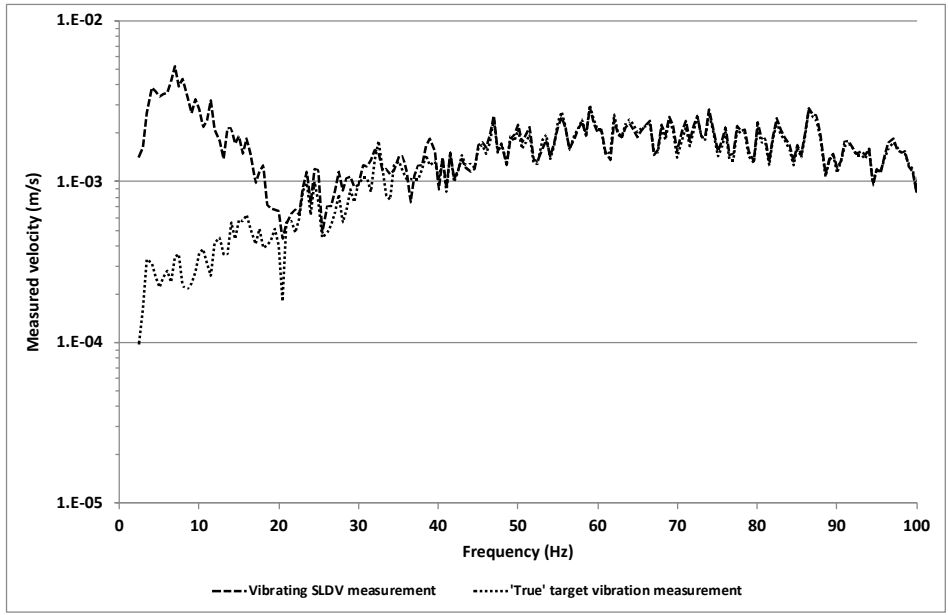

b

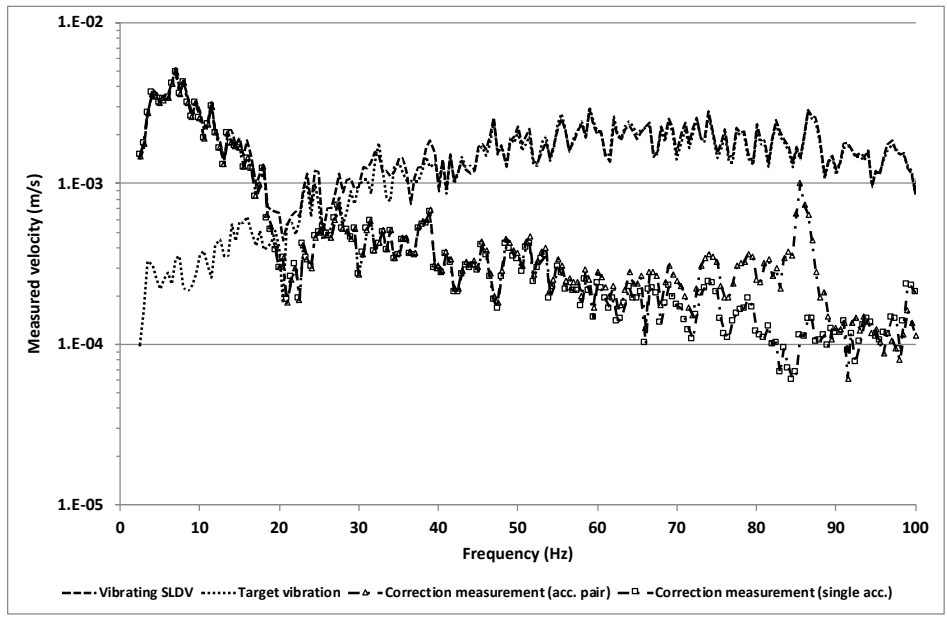

c

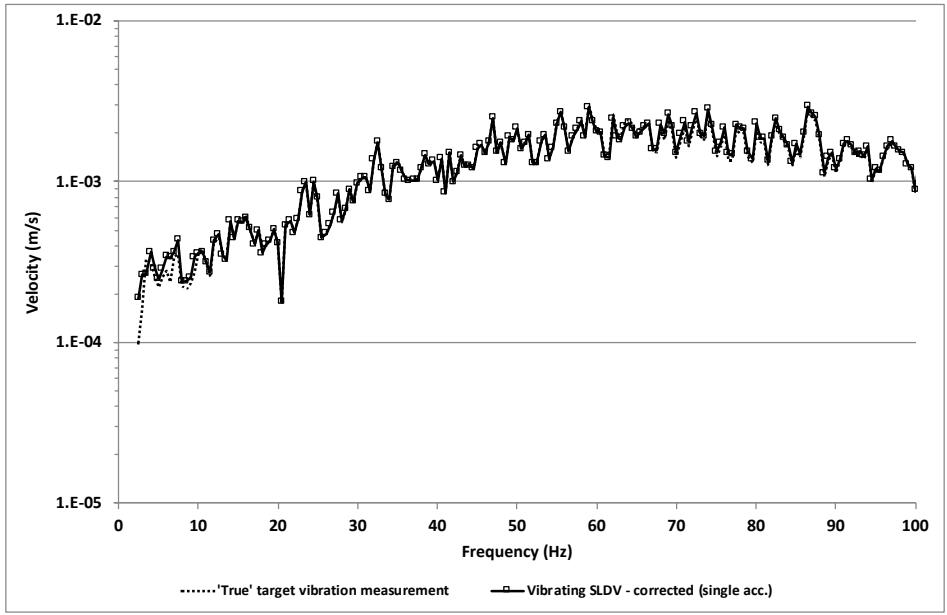

d

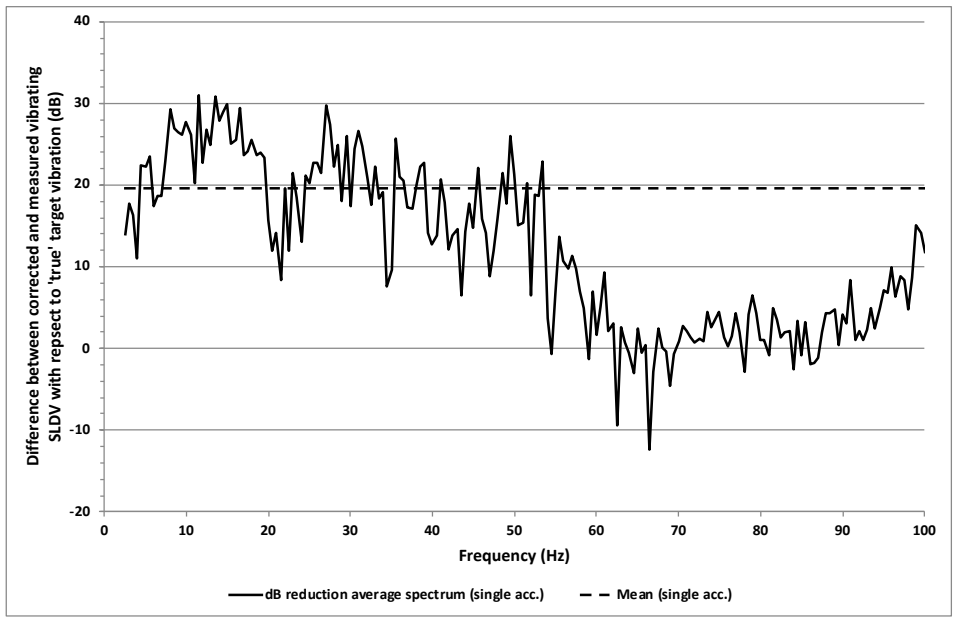

e

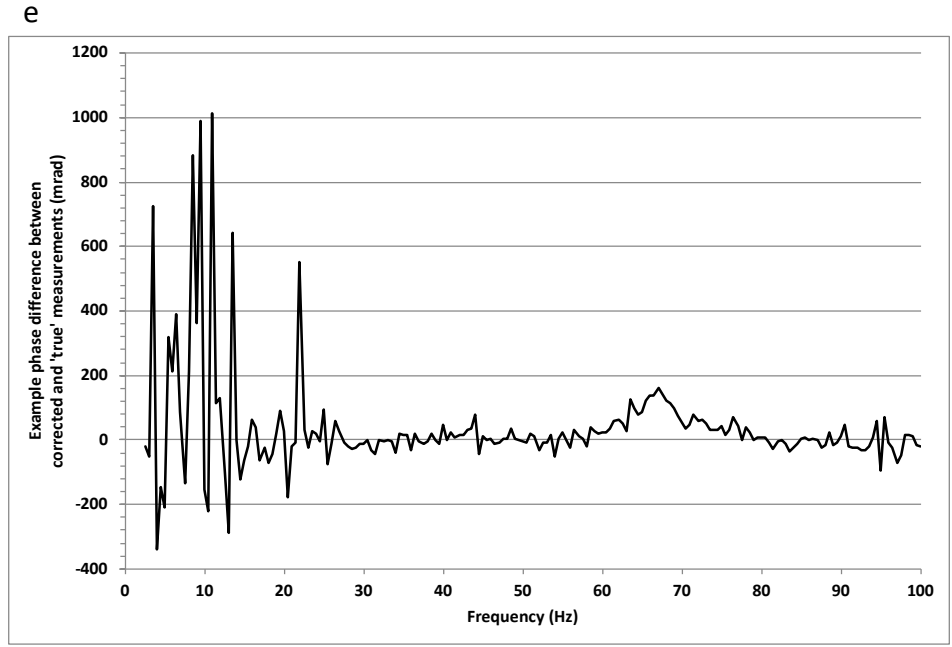

Figure 5. For simultaneous target *and* SLDV sensor head vibration, comparison between measurements from the SLDV and of the 'true' target vibration; a) measured averaged spectra, b) proposed correction measurements, c) corrected averaged spectrum (single accelerometer only), d) averaged dB reduction (single accelerometer only) and e) example phase difference spectrum (single accelerometer only).

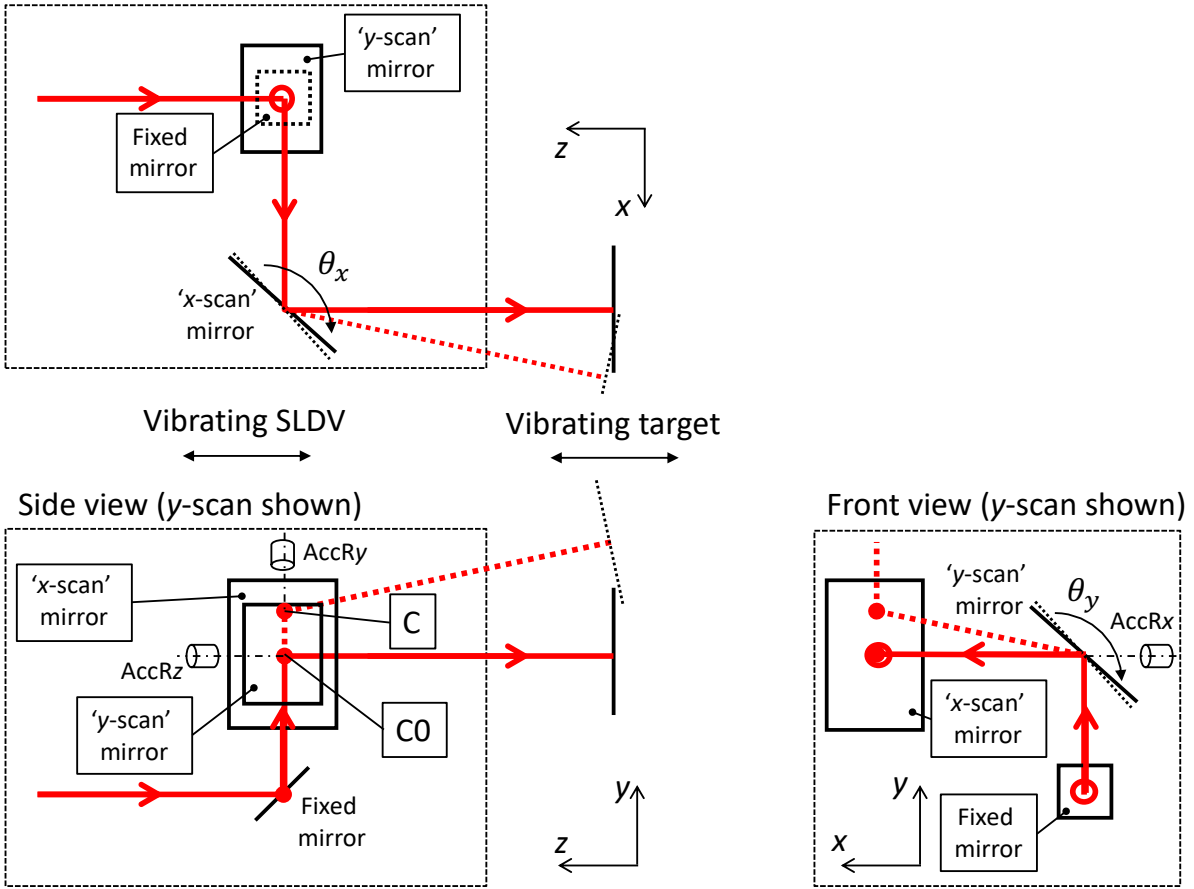

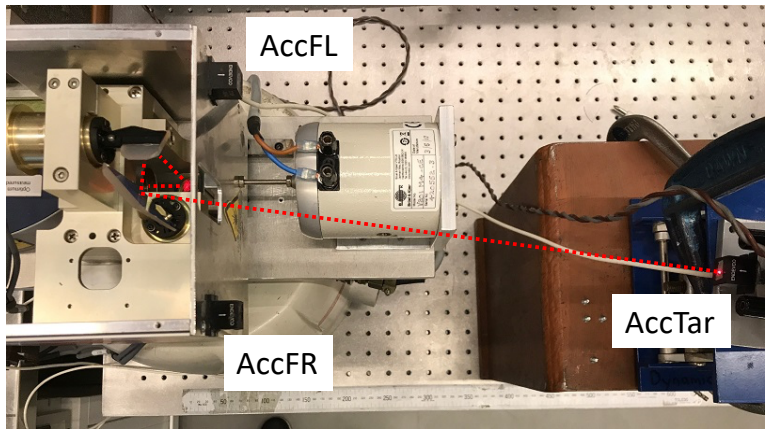

Figure 6. (a) Top, side and front view schematic diagrams showing laser beam path and correction accelerometer locations. (b) Physical set-up showing mirror rotations, beam path (super-imposed) and target.

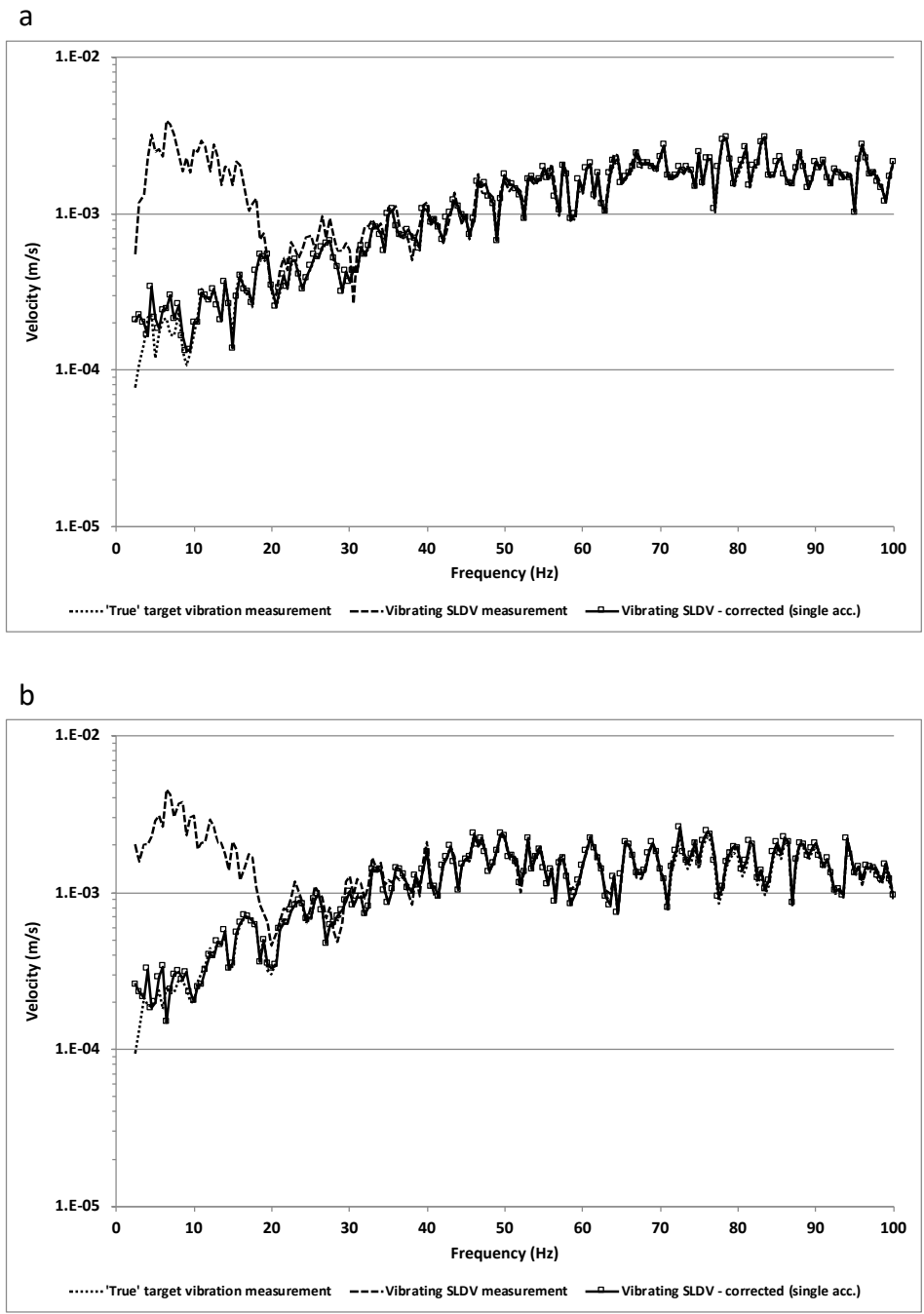

Figure 7. Comparison between measurements from the vibrating SLDV and the 'true' target vibration *during scanning* for single accelerometer correction without geometrical weighting; a) -4° and b) +12°.

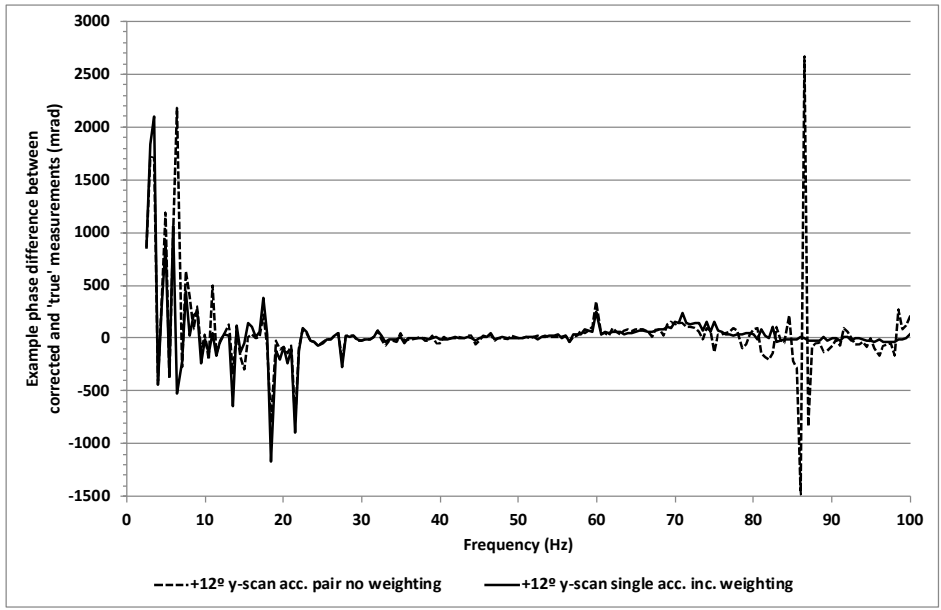

Figure 8. Example phase difference plot comparing correction options *during scanning*.